\documentclass[reprint,superscriptaddress,amsmath,amssymb,twocolumn,prl]{revtex4-2}

\usepackage{amsmath}
\usepackage{mathtools}
\usepackage{amssymb}
\usepackage{graphicx}
\usepackage{bm}
\usepackage[colorlinks, linkcolor=blue, citecolor=blue, urlcolor=blue, breaklinks=true]{hyperref}
\usepackage{color,dsfont,multirow,booktabs}
\usepackage{braket}
\usepackage{tabularx}
\usepackage{lipsum,booktabs}
\usepackage{subfiles}
\usepackage{subcaption}
\usepackage{caption}
\usepackage[dvipsnames]{xcolor}
\usepackage{ragged2e}
\usepackage{orcidlink}
\usepackage[T1]{fontenc}
\usepackage[normalem]{ulem}

\newcommand{\commentold}[1]{}
\DeclareMathSymbol{:}{\mathpunct}{operators}{"3A}

\newcommand{\figpanel}[2]{\hyperref[#1]{\ref*{#1}(#2)}}

\begin{document}
\date{\today}

\title{Quantum Batteries as Work Sources for Phase-Locked Parametric Amplification}

\author{Borhan Ahmadi\orcidlink{0000-0002-2787-9321}}
\email{borhan.ahmadi@ug.edu.pl}
\address{International Centre for Theory of Quantum Technologies, University of Gdańsk, ul. prof. Marii Janion 4, 80-309 Gdańsk, Poland}

\begin{abstract}
Quantum batteries have been proposed as locally precharged work sources for superconducting quantum technologies, suggesting a route to reduce continuously supplied microwave drives. Here we ask whether the pump tone of a quantum-limited parametric amplifier can be replaced, or strongly duty-cycled, by a finite bosonic quantum battery. Quantizing the pump of a nondegenerate parametric amplifier exposes a resource distinction hidden in the classical description: stored pump energy can generate signal-idler photons, but pump phase coherence is required to generate a phase-locked amplifier field. In a closed trilinear model, coherent and phase-randomized coherent pumps with the same photon-number distribution produce comparable pair numbers, yet only the coherent pump produces anomalous two-mode coherence and an EPR-squeezed interference dip. Including leakage, we collect the emitted fields into cascaded temporal modes. At matched collector bandwidth, the coherent pump gives \(I_{\min}^{(f)}=0.553\), whereas the phase-randomized pump gives \(I_{\min}^{(f)}=1.94\) at nearly identical collected energy. Weak amplitude squeezing slightly improves the dip by reducing finite-pump number fluctuations while preserving the coherent displacement. Thus battery-powered parametric amplification requires phase-coherent stored energy, possibly assisted by number-noise reduction, rather than stored energy alone.
\end{abstract}

\maketitle
\textit{Introduction}---
Quantum batteries (QBs) were introduced as finite quantum systems whose stored energy and extractable work can be enhanced or controlled using genuinely quantum resources. Early work showed that entanglement can increase the extractable work from ensembles of quantum systems \cite{AlickiFannes2013}, and later studies established that collective operations can enhance charging power beyond independent parallel charging \cite{Binder2015,Campaioli2017}. This motivated concrete many-body and light-matter implementations, including Dicke-type solid-state quantum batteries and charger-mediated energy-transfer models \cite{Ferraro2018,PhysRevA.107.042419,PhysRevApplied.23.024010,Andolina2018,Farina2019,Razzoli_2025}. The field has since expanded to open-system stabilization, feedback and measurement-assisted charging, optimal-control and reinforcement-learning protocols, and experimental demonstrations of collective superabsorption in organic microcavities \cite{adma.202415073,qhz8-mvfb,malavazi2025charge,Gherardini2020,zakavati2025optimizing,Mitchison2021,PhysRevA.109.042411,lu2021optimal,ahmadi2025harnessing,kamin2023steady,Quach2022,yang2024three,Erdman2024,bv4w-jr6q,ahmadi2026ChiralSqueezing}. Recent reviews emphasize that the central question is not merely how much energy is stored, but how quantum resources such as coherence, correlations, collective coupling, and ergotropy make that energy operationally useful \cite{Campaioli2024,ferraro2026opportunities,ahmadi2026LandauZener}. This shift raises a natural but largely unexplored question: can a quantum battery be used as the work source of a functional quantum technology, and if so, which part of its stored energy is actually useful for the task?

This question has recently acquired a concrete hardware dimension in superconducting platforms. Kurman \textit{et al.} proposed a shared resonator quantum battery for superconducting quantum computation, where qubit gates are implemented by tuning qubit frequencies relative to a precharged bosonic mode rather than by applying individual microwave drive tones; in their heat-load estimate, removing those drive lines can reduce cryogenic wiring overhead and, with superconducting wiring, increase the cooling-limited qubit capacity by roughly a factor of four \cite{Kurman2026}. A closely related bottleneck appears in superconducting readout chains. Cryogenic microwave cabling and attenuation generate both passive and active heat loads \cite{Krinner2019}, while quantum-limited parametric amplifiers normally require strong microwave pump tones. Recent traveling-wave parametric-amplifier work identifies off-chip pump-routing components as sources of loss, complexity, and limited scalability \cite{Denney2026}; pump-efficient Josephson-parametric-amplifier work links low pump-added efficiency to cryostat heat load and limits on the number of devices that can be hosted in a dilution refrigerator \cite{Hougland2025}; and dc-powered quantum-limited amplification has been proposed specifically to eliminate microwave pump-tone infrastructure \cite{Nehra2026}. These developments motivate the amplifier-specific problem addressed here: whether a locally precharged pump mode can replace, or at least strongly duty-cycle, the continuously supplied microwave pump of a quantum-limited amplifier.

Parametric amplification provides a stringent test of this idea because the pump is not merely an energy source; it is also the phase reference that locks the emitted signal-idler field. Phase-preserving amplifiers are normally modeled in the undepleted-pump approximation, where the pump is an ideal classical phase reference and the signal-idler dynamics is generated by a two-mode-squeezing Hamiltonian. This assumption underlies the standard theory of quantum-limited amplification and its superconducting-circuit implementations \cite{Caves1982,Clerk2010,Bergeal2010}. At the same time, fully quantum trilinear Hamiltonians with quantized pump, signal, and idler modes have long been studied in connection with quantum-mechanical amplification, frequency conversion, field statistics, entanglement, pump depletion, and nonclassical effects \cite{WallsBarakat1970,DrobnyJex1992,DrobnyJex1993,BandillaDrobnyJex1996,FerneeKinslerDrummond1995,XingRalph2023,ChinniEtAl2024}. Pump coherence is also known to affect down-conversion correlations, including spatial-coherence effects in spontaneous parametric down-conversion and phase effects in coherently stimulated down-conversion with a quantized pump \cite{GieseEtAl2018,BirrittellaAlsingGerry2020}. Here we ask what resource of a finite bosonic pump battery is certified by phase-locked amplification. By comparing coherent, phase-randomized, Fock and squeezed displaced pump batteries with the same photon-number distribution, and by measuring the emitted radiation in explicit cascaded collector modes, we show that output energy transfer and phase-sensitive amplification certify different resources. Stored pump energy can generate signal-idler photons, but phase-coherent pump-battery energy is required to generate a phase-locked amplifier field.

The goal is therefore not to claim that every classical control or calibration line can be removed. Rather, it is to identify when a precharged on-chip quantum mode can take over the phase-coherent work-source role of a continuously supplied microwave pump, and what energetic and coherence resources such a replacement would require.

\begin{figure}
    \centering
    \includegraphics[width=\linewidth]{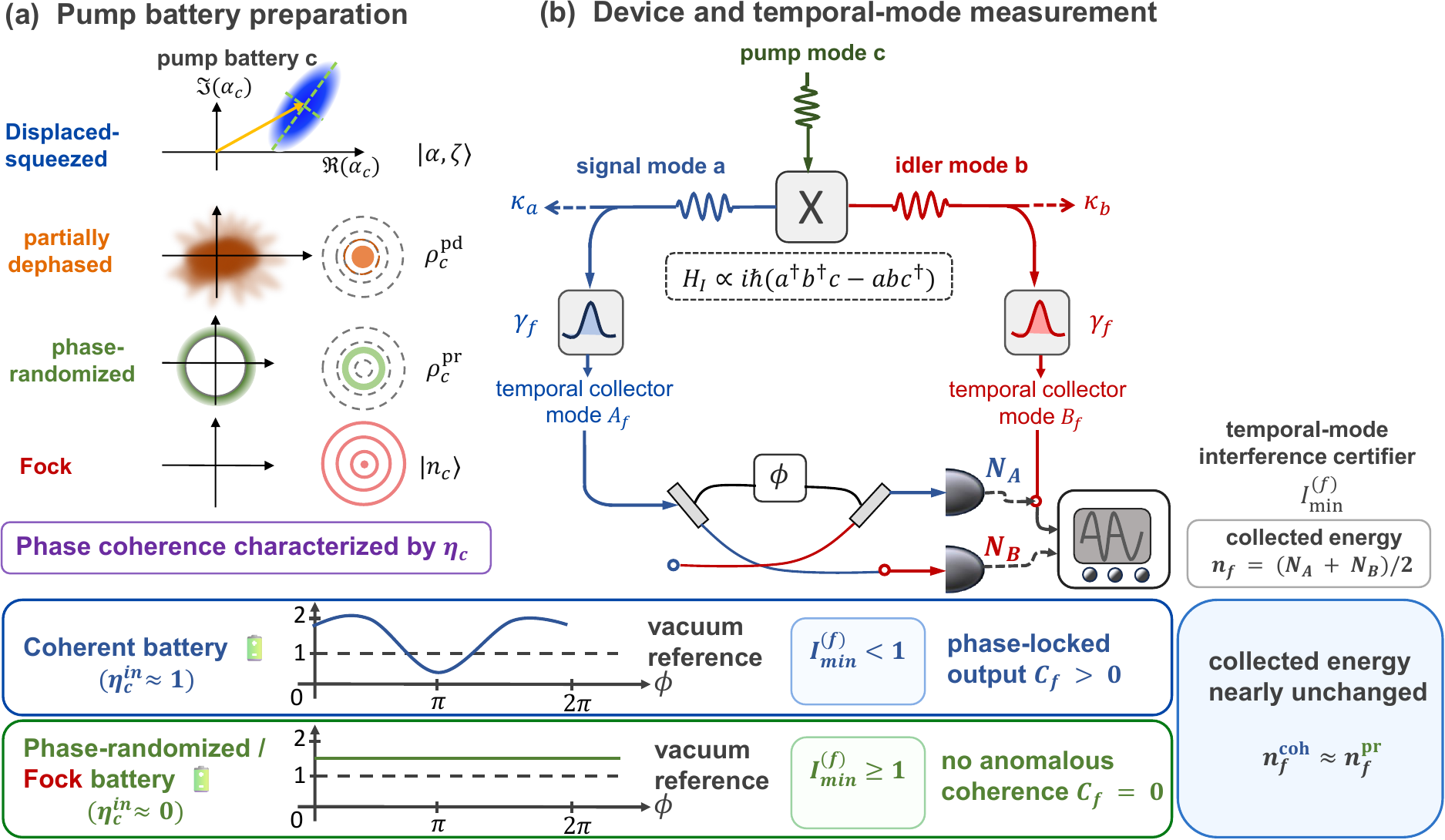}
    \caption{Phase-coherent quantum battery powering a parametric amplifier.
    (a) A finite bosonic QB (mode \(c\)) is prepared with fixed mean energy \(\bar n_c\) and fixed nominal squeezing rate \(g\sqrt{\bar n_c}=\lambda\), while its phase coherence is varied across coherent, displaced-squeezed, partially dephased, phase-randomized, and number-state-like preparations. The coherent fraction \(\eta_c^{\rm in}=|\langle c\rangle|^2/\langle c^\dagger c\rangle\) quantifies the battery phase reference.
    (b) The pump battery drives a nondegenerate three-wave-mixing amplifier. Signal and idler photons emitted through ports \(\kappa_a\) and \(\kappa_b\) are collected into temporal modes \(A_f\) and \(B_f\), which are then interfered to measure the phase-sensitive certifier \(I_{\min}^{(f)}\) and the collected energy \(n_f=(N_A+N_B)/2\).\justifying}\label{Schematic}
\end{figure}

The physical setting and operational logic are summarized in Fig.~\ref{Schematic}. A finite bosonic pump battery with fixed mean energy and fixed nominal classical squeezing rate powers a nondegenerate parametric amplifier, while its quantum state and phase coherence are varied. The emitted signal and idler fields are collected into temporal modes and interfered. The key comparison is then between output energy transfer and phase-sensitive temporal-mode coherence: two pump batteries can deposit nearly the same collected energy, but only the phase-coherent battery produces a sub-vacuum interference dip.

\textit{Model}.--
We consider a nondegenerate parametric amplifier in which the pump is treated as a finite quantum system rather than as an externally prescribed classical tone. The signal, idler, and pump-battery modes are denoted by \(a\), \(b\), and \(c\), respectively. In the resonant interaction picture the closed trilinear Hamiltonian is
\begin{equation}
H_I
=
i\hbar g
\left(
c a^\dagger b^\dagger
-
c^\dagger a b
\right).
\label{eq:trilinear_H}
\end{equation}
The classical-pump amplifier is recovered by replacing the pump operator by a complex amplitude,
\(
c\rightarrow \beta e^{-i\phi_p}
\)
and
\(
\lambda=g|\beta|,
\)
which gives the usual two-mode-squeezing interaction
\(
H_{\rm cl}
=
i\hbar\lambda
\left(
e^{-i\phi_p}a^\dagger b^\dagger
-
e^{i\phi_p}ab
\right).
\)
Thus the standard amplifier limit hides the physical state of the work source in the classical number \(\beta e^{-i\phi_p}\). Our goal is to resolve this source as a finite quantum battery and to determine which battery resource is certified by amplification.

We first calibrate all pump-battery states by the same nominal classical squeezing rate,
\(
g\sqrt{\bar n_c}=\lambda,
\)
where \(\bar n_c=\langle c^\dagger c\rangle\) is the initial battery energy in photon units. This calibration fixes the classical operating point while allowing the quantum state of the pump battery to vary. We compare coherent, Fock, and phase-randomized coherent pump states. The latter has the same photon-number distribution as a coherent state but no first-order phase coherence,
\begin{equation}
\rho_{\rm pr}
=
\sum_{n=0}^\infty
e^{-\bar n_c}
\frac{\bar n_c^n}{n!}
|n\rangle\langle n|.
\end{equation}
We also use the interpolating family
\begin{equation}
\rho_c(0)
=
w|\alpha\rangle\langle\alpha|
+
(1-w)\rho_{\rm pr},
\quad
|\alpha|^2=\bar n_c,
\label{eq:partial_dephase_family}
\end{equation}
for which the initial coherent fraction is
\(
\eta_c^{\rm in}
= |\langle c\rangle|^2/\langle c^\dagger c\rangle = w^2.
\)
The partially dephased state interpolates between a coherent pump and a fully phase-randomized coherent pump while keeping the same Poissonian photon-number distribution. It reduces the first-order phase reference according to \(\eta_c^{\rm in}=w^2\), without changing the stored pump energy. In the closed trilinear model, the relevant diagnostic is not only the generated pair number,
\(
n_{\rm pair} = \left(\langle a^\dagger a\rangle+\langle b^\dagger b\rangle\right)/2,
\)
but also the anomalous two-mode coherence \(M = \langle ab\rangle.\) We quantify this phase-sensitive coherence by the normalized amplifier coherence
\begin{equation}
C_{\rm amp} = \frac{|M|}
{\sqrt{n_{\rm pair}(n_{\rm pair}+1)}}.
\label{eq:Camp}
\end{equation}
This quantity equals unity for an ideal two-mode squeezed vacuum. The corresponding EPR quadrature certifier is
\(
V_-
=
{\rm Var}(X_a-X_b)
=
\langle a^\dagger a\rangle+\langle b^\dagger b\rangle+1
-
2{\rm Re}\langle ab\rangle,
\)
with \(X_j=(j+j^\dagger)/\sqrt 2\). In this convention the vacuum level is \(V_-=1\).

These diagnostics separate energy transfer from phase-coherent amplification. A phase-randomized coherent pump produces essentially the same pair number as a coherent pump with the same \(\bar n_c\), because both have the same photon-number distribution. However, it gives \(\langle ab\rangle=0\), and therefore it cannot produce an EPR-squeezed, phase-locked amplifier output. Likewise, a Fock pump contains sharply defined energy but no phase reference, so it can generate pairs while failing the two-mode coherence test. Thus pair generation alone does not certify the pump battery as a coherent work source.

This distinction is directly measurable by the two-mode interference observable
\(
I(\phi) = \left\langle
\left(a+e^{i\phi}b^\dagger\right)^\dagger
\left(a+e^{i\phi}b^\dagger\right)
\right\rangle.
\)
Writing \(n_a=\langle a^\dagger a\rangle\) and \(n_b=\langle b^\dagger b\rangle\), the bosonic commutation relations give
\(
I(\phi)
=
n_a+n_b+1
+
2{\rm Re}
\left(
e^{-i\phi}\langle ab\rangle
\right),
\)
leading to
\(
I_{\min}
=
n_a+n_b+1-2|\langle ab\rangle|,
\)
\(
I_{\max}
=
n_a+n_b+1+2|\langle ab\rangle|.
\)
The interference visibility, defined by the standard fringe contrast \cite{MandelWolf1995}, is
\begin{equation}
{\cal V}_{\rm int}
=
\frac{I_{\max}-I_{\min}}{I_{\max}+I_{\min}}
=
\frac{2|\langle ab\rangle|}{n_a+n_b+1}.
\label{eq:visibility}
\end{equation}
For the family in Eq.~\eqref{eq:partial_dephase_family}, dephasing leaves the generated pair number unchanged but reduces the anomalous coherence as
\(
\langle ab\rangle
=
\sqrt{\eta_c^{\rm in}}\,
\langle ab\rangle_{\rm coh}.
\)
Consequently the sub-vacuum interference condition \(I_{\min}<1\) requires
\begin{equation}
\eta_c^{\rm in}
>
\left(
\frac{n_{\rm pair}}{|\langle ab\rangle_{\rm coh}|}
\right)^2.
\label{eq:closed_threshold_general}
\end{equation}
In the stiff-pump limit we have
\(
n_{\rm pair}=\sinh^2\tau
\)
and
\(
|\langle ab\rangle_{\rm coh}|=\sinh\tau\cosh\tau,
\)
so the closed pulsed model gives the simple guide
\(
\eta_{c,\rm crit}^{\rm in}
\simeq
\tanh^2\tau
\)
(see Note~1 of the SM \cite{SuppMat}). This relation shows that higher-gain operation requires a larger phase-coherent fraction of the pump battery.

The closed model therefore identifies the resource distinction in its simplest form: stored pump energy controls pair generation, while pump phase coherence controls whether the generated pairs form a phase-locked amplifier field. We now ask whether the same distinction survives in the experimentally relevant situation where the signal and idler leak out of the resonator and are measured as temporal output modes.

\textit{Open-system temporal-mode certification}---
We next include leakage through the signal and idler ports, as in a JPC-like amplifier. Using the standard Markovian input-output description of damped bosonic modes \cite{Lindblad1976,GardinerCollett1985,Clerk2010}, the source modes obey
\(
\dot\rho
=
-\frac{i}{\hbar}[H_I,\rho]
+
\kappa_a{\cal D}[a]\rho
+
\kappa_b{\cal D}[b]\rho
+
\kappa_c{\cal D}[c]\rho ,
\)
where
\(
{\cal D}[o]\rho
=
o\rho o^\dagger
-
\frac{1}{2}
\left\{
o^\dagger o,\rho
\right\}.
\)
The corresponding input-output relations are
\(
a_{\rm out}=a_{\rm in}+\sqrt{\kappa_a}a
\)
and
\(
b_{\rm out}=b_{\rm in}+\sqrt{\kappa_b}b
\)
up to the usual convention-dependent sign of the output field \cite{GardinerCollett1985,Clerk2010}.

\begin{figure}
    \centering
    \includegraphics[width=1\linewidth]{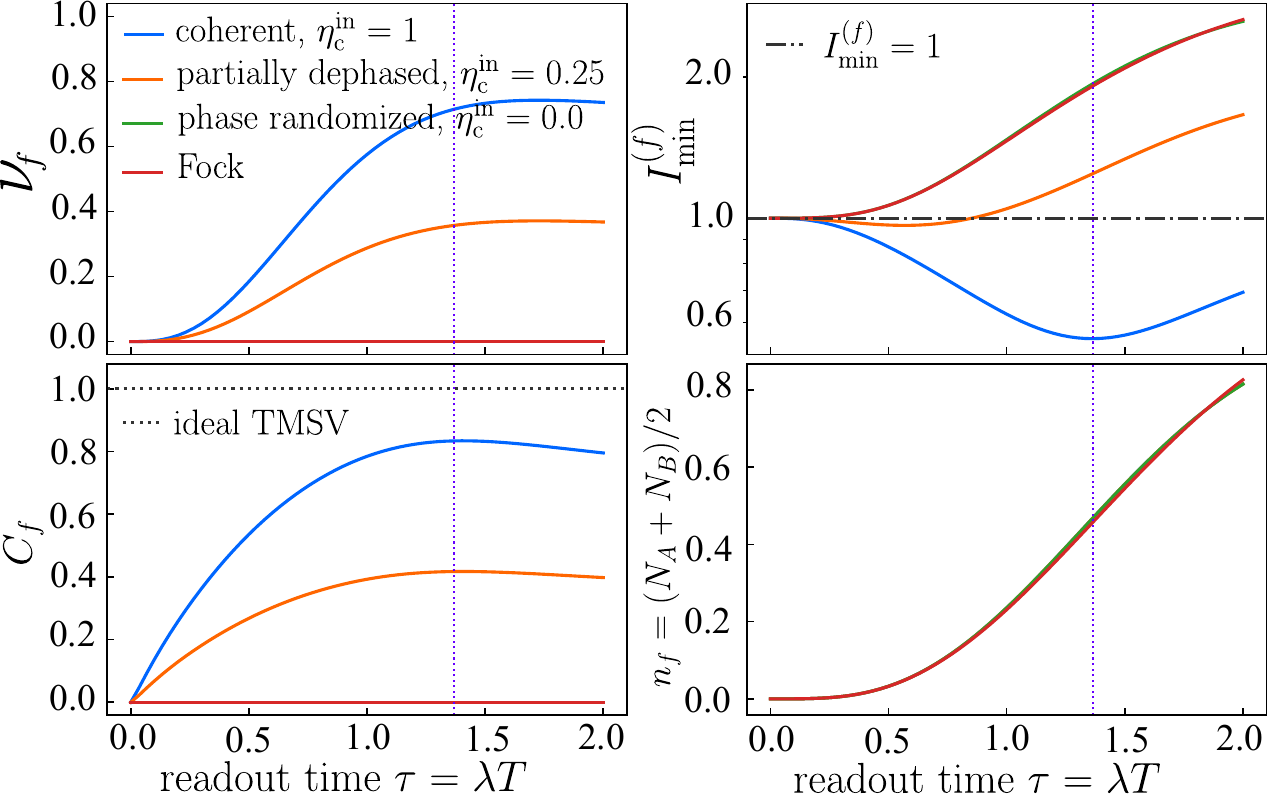}
    \caption{Collector-mode certification of finite-battery parametric amplification. The leaking signal and idler fields are collected into auxiliary temporal modes \(A\) and \(B\) with matched bandwidth \(\gamma_f/\lambda=2\). The panels show the collected phase-locking visibility \({\cal V}_f\), the interference certifier \(I_{\min}^{(f)}\), the normalized collected pair coherence \(C_f\), and the collected pair number \(n_f=(N_A+N_B)/2\). A coherent pump battery produces finite temporal-mode phase locking, \(C_f>0\), and a sub-vacuum interference dip \(I_{\min}^{(f)}<1\). Phase-randomized coherent and Fock pumps collect comparable energy but have no anomalous temporal-mode coherence and remain above the interference threshold. The dotted vertical purple line marks the optimal coherent-pump readout time \(\tau_\star=1.37\).
\justifying}\label{fig:collector_basic}
\end{figure}

To avoid relying on equal-time output proxies, we explicitly collect the emitted fields into auxiliary bosonic modes \(A\) and \(B\). This is a cascaded-systems construction: the same composite-jump-operator and cascaded-Hamiltonian structure is used, for example, to inject an itinerant microwave photon from a source mode into an absorber mode in Ref.~\cite{Royer2018}, following the general cascaded-systems formalism of Refs.~\cite{Gardiner1993,Carmichael1993}. For each output channel we take
\(
L_a=\sqrt{\kappa_a}a+\sqrt{\gamma_A}A
\)
and
\(
L_b=\sqrt{\kappa_b}b+\sqrt{\gamma_B}B
\),
together with the cascaded Hamiltonian
\(
H_{\rm cas} = \frac{i\hbar}{2}\left(\sqrt{\kappa_a\gamma_A}
\left(a^\dagger A - A^\dagger a\right) + \sqrt{\kappa_b\gamma_B}
\left(b^\dagger B - B^\dagger b\right)\right)
\)
The full collector-mode master equation is then
\(
\dot\rho
=
-\frac{i}{\hbar}
\left[
H_I+H_{\rm cas},\rho
\right]
+
{\cal D}[L_a]\rho
+
{\cal D}[L_b]\rho
+
\kappa_c{\cal D}[c]\rho .
\)
The collectors are initially in the vacuum. Their moments at readout time \(T\),
\(
N_A(T)=\langle A^\dagger A\rangle_T,
\)
\(
N_B(T)=\langle B^\dagger B\rangle_T,
\)
\(
M_{AB}(T)=\langle AB\rangle_T,
\)
define a canonical temporal-mode output witness. The collected pair number is
\(
n_f(T)=
\left(N_A(T)+N_B(T)\right)/2,
\)
and the normalized collected pair coherence is
\(
C_f(T)
= |M_{AB}(T)|/\sqrt{n_f(T)\left[n_f(T)+1\right]} .
\)
The phase-sensitive collector interference observable is
\begin{equation}
I_f(\phi,T)
=
\left\langle
\left(A+e^{i\phi}B^\dagger\right)^\dagger
\left(A+e^{i\phi}B^\dagger\right)
\right\rangle_T.
\end{equation}
This observable is our temporal-mode version of the standard phase-sensitive two-mode correlation measured in nondegenerate parametric amplification \cite{Caves1982,Clerk2010}. Its algebraic form follows directly from the bosonic commutator \([B,B^\dagger]=1\):
\begin{equation}
I_f(\phi,T)
=
N_A(T)+N_B(T)+1
+
2{\rm Re}
\left[
e^{-i\phi}M_{AB}(T)
\right],
\end{equation}
thus
\(
I_{\min}^{(f)}(T) = N_A(T) + N_B(T) + 1 - 2|M_{AB}(T)| .
\)
The corresponding collector visibility is
\(
{\cal V}_f=(I_{\max}^{(f)}-I_{\min}^{(f)})/(I_{\max}^{(f)}+I_{\min}^{(f)})
\),
with
\(
I_{\max}^{(f)}(T)=N_A(T)+N_B(T)+1+2|M_{AB}(T)|.
\)
The vacuum reference is \(I_{\min}^{(f)}=1\). Therefore \(I_{\min}^{(f)}<1\) certifies sub-vacuum phase-sensitive interference in the collected temporal modes. This collector-mode witness gives the central open-system test of the battery-resource distinction. We use matched collectors, \(\gamma_A=\gamma_B=\gamma_f\), and first choose \(\gamma_f/\lambda=2\), which is close to the best matched bandwidth identified below. The resulting temporal-mode dynamics are shown in Fig.~\ref{fig:collector_basic}. The four panels should be read together. The collected pair number \(n_f\) grows almost independently of whether the pump is coherent or phase randomized, showing that dephasing the pump does not strongly suppress energy transfer into the measured output mode. In contrast, the phase-sensitive quantities respond sharply to pump coherence: the coherent pump produces finite phase locking \({\cal V}_f\), large normalized pair coherence \(C_f\), and a sub-vacuum dip \(I_{\min}^{(f)}<1\), whereas the phase-randomized and Fock pumps have \(C_f=0\) and remain above the vacuum reference. Thus Fig.~\ref{fig:collector_basic} already separates two operational effects: emission of output energy and generation of a phase-locked amplifier field.

For the main numerical example we choose a stable below-threshold amplifier regime with reduced pump strength
\begin{equation}
\xi = \frac{2\lambda}{\sqrt{\kappa_a\kappa_b}} = 0.8,
\quad
\bar n_c=5,
\end{equation}
where \(\xi\) is the stiff-pump gain parameter, while \(\bar n_c\) fixes the finite pump-battery size. In this regime, the coherent pump reaches its optimal collector readout at
\(
\tau_\star=\lambda T_\star=1.37.
\)
At this readout time, Fig.~\ref{fig:collector_basic} gives
\(
I_{\min}^{(f)}(\tau_\star)=0.553,
\)
\(
C_f(\tau_\star)=0.834,
\)
\(
n_f(\tau_\star)=0.4698
\)
for the coherent pump. In contrast, a phase-randomized coherent pump with the same photon-number distribution gives
\(
I_{\min}^{(f)}(\tau_\star)=1.94,
\)
\(
C_f(\tau_\star)=0,
\)
\(
n_f(\tau_\star)=0.4692 .
\)
The collected energies are therefore nearly identical, while the phase-sensitive interference is present only for the coherent pump. A Fock pump similarly gives \(C_f=0\) and remains above the interference threshold. The distinction is therefore not between emitting and not emitting radiation; it is between emitting incoherent radiation and emitting a phase-locked amplifier field.
\begin{figure}
    \centering
    \includegraphics[width=0.95\linewidth]{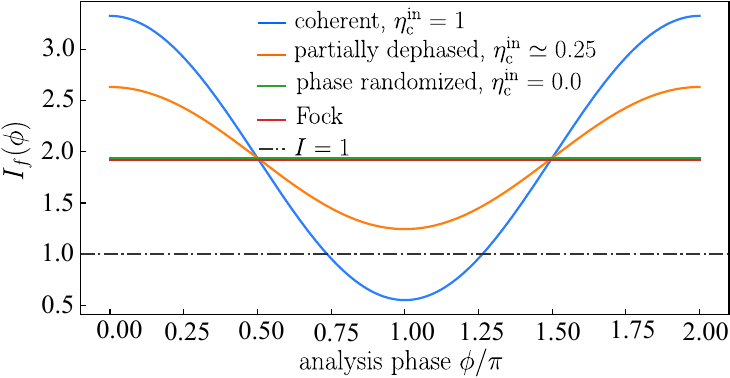}
    \caption{Collected temporal-mode interference fringe at the optimal readout time. The interference observable \(I_f(\phi)\) is evaluated from the collector modes \(A\) and \(B\) at \(\tau_\star=1.37\). The coherent pump gives a high-contrast fringe whose minimum is below the vacuum reference \(I=1\), with \(I_{\min}^{(f)}=0.553\). The phase-randomized coherent pump has nearly the same collected energy but produces a flat trace above the vacuum reference because \(M_{AB}=0\). The Fock pump also lacks a phase reference and therefore does not generate a phase-locked amplifier field.
\justifying}\label{fig:collector_fringe}
\end{figure}

The same conclusion is seen more directly in the phase-resolved collector fringe in Fig.~\ref{fig:collector_fringe}. At \(\tau_\star\), the coherent pump produces a high-contrast interference pattern whose minimum lies below the vacuum reference \(I=1\). This is the temporal-mode signature of phase-sensitive amplification. The phase-randomized coherent pump, despite depositing essentially the same collected energy, produces a flat trace above the vacuum reference because \(M_{AB}=0\). The Fock pump is also flat for the same reason: it contains stored energy but no phase reference. Thus the fringe measurement makes the resource separation operational. Pair production alone is not enough; only phase-coherent battery energy generates a phase-locked output field.

The Hilbert-space truncation is controlled by monitoring source- and collector-edge populations at the physical readout time; the corresponding diagnostics and a complementary larger-source-cutoff check are given in Note~2 of the SM \cite{SuppMat}.

\begin{figure}
    \centering
    \includegraphics[width=1\linewidth]{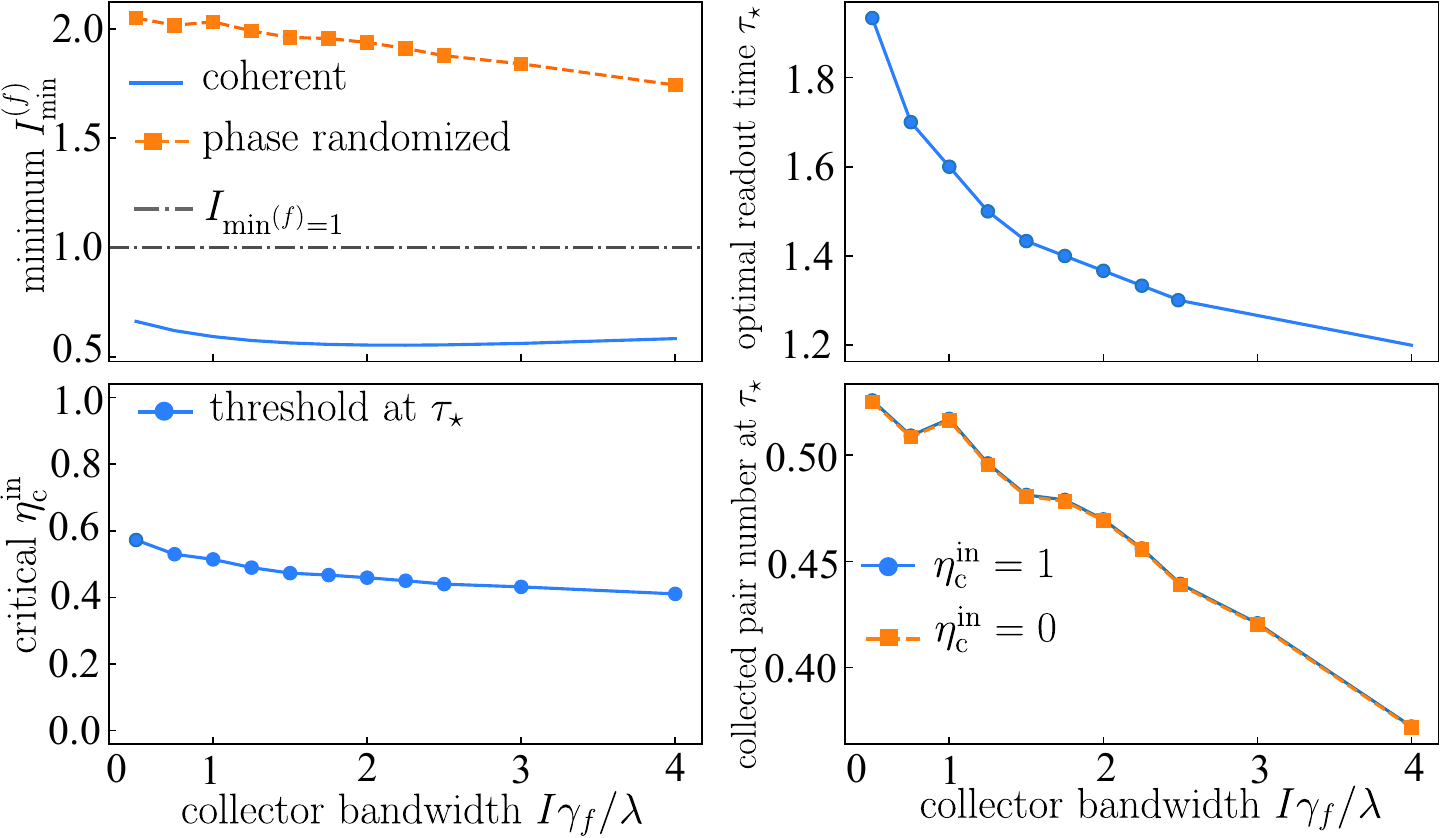}
    \caption{Collector-bandwidth dependence of the temporal-mode certifier. The matched collector bandwidth \(\gamma_f=\gamma_A=\gamma_B\) is varied at fixed \(\bar n_c=5\), \(\xi=2\lambda/\sqrt{\kappa_a\kappa_b}=0.8\), and \(\lambda T_{\max}=2\). The upper-left panel shows the minimum temporal-mode interference dip obtained by optimizing over the readout time. The coherent pump remains below the vacuum reference \(I_{\min}^{(f)}=1\), whereas the phase-randomized pump remains above it. The coherent-pump dip improves as the bandwidth is increased up to \(\gamma_f/\lambda\simeq2\text{--}2.25\), and then becomes weaker for broader collectors. The upper-right panel shows the corresponding optimal readout time \(\tau_\star\), which decreases with increasing bandwidth. The lower-left panel gives the coherent-fraction threshold extracted at \(\tau_\star\). The lower-right panel shows that coherent and phase-randomized pumps deposit nearly identical collected pair numbers at the same readout time. We use \(\gamma_f/\lambda=2\) as a matched-bandwidth working point because it lies on the optimum plateau and differs from the best scanned point, \(\gamma_f/\lambda=2.25\), only at the \(10^{-3}\) level in \(I_{\min}^{(f)}\).
    \justifying}
\label{fig:gamma_scan}
\end{figure}

\textit{Matched temporal mode and bandwidth dependence}.--
The collector bandwidth determines which temporal mode of the emitted field is measured. We therefore scan \(\gamma_f/\lambda\) and, for each bandwidth, choose the coherent-pump readout time \(\tau_\star(\gamma_f)=\arg\min_\tau I_{\min}^{(f)}(\tau)\). The extended scan is summarized in Fig.~\ref{fig:gamma_scan}. As the bandwidth is increased from narrow values, the coherent-pump interference dip deepens and the optimal readout time shifts to earlier times. The improvement then saturates around \(\gamma_f/\lambda\simeq2\text{--}2.25\), after which broader collectors give a weaker dip. The best point in the scan is \(\gamma_f/\lambda=2.25\), with \(I_{\min}^{(f)}=0.5527\), but this is almost indistinguishable from \(I_{\min}^{(f)}=0.5533\) at \(\gamma_f/\lambda=2\). We therefore use \(\gamma_f/\lambda=2\) as a simple matched-bandwidth working point in the main calculations. Across the whole scan, the coherent and phase-randomized pumps deposit nearly identical collected energy at the matched readout time, while only the coherent pump remains below the vacuum interference reference. Thus Fig.~\ref{fig:gamma_scan} shows that the resource separation is not an artifact of a single detection window: a properly matched temporal output mode reveals phase-sensitive amplification only when the pump battery contains phase coherence.

Finally, we ask whether nonclassical pump-battery engineering can improve the amplifier task. In Note~3 of the SM \cite{SuppMat}, we analyze displaced squeezed pump batteries,
\(
|\alpha,\zeta\rangle_c
=
D_c(\alpha)S_c(\zeta)|0\rangle,
\)
\(
\zeta=re^{i\vartheta_s},
\)
at fixed stored energy
\(
\bar n_c=|\alpha|^2+\sinh^2 r.
\)
At fixed \(\bar n_c\), squeezing therefore introduces a direct tradeoff: the coherent displacement that supplies the first-order pump phase reference is reduced according to
\(
\eta_c^{\rm in}
=
\frac{|\langle c\rangle|^2}{\langle c^\dagger c\rangle}
=
1-\frac{\sinh^2 r}{\bar n_c}.
\)
Amplitude squeezing can nevertheless be useful because, for weak \(r\), it reduces pump-number fluctuations while leaving most of the coherent displacement intact. The cascaded-collector calculation shows a small but systematic operational advantage in this regime. For \(\bar n_c=2\), the optimized temporal-mode dip improves from \(I_{\min}^{(f)}=0.526\) for a coherent pump to \(I_{\min}^{(f)}=0.516\) for a weakly amplitude-squeezed pump. At the main parameter set used for the collector-mode figures, \(\xi=0.8\), \(\gamma_f/\lambda=2\), and \(\bar n_c=5\), the corresponding improvement is \(0.553\to0.547\). In both cases the improvement occurs while the coherent fraction remains very close to unity, whereas the normalized pump-number variance is reduced by about \(25\%\). Stronger squeezing reverses this gain because it transfers energy out of the coherent displacement and reduces the phase-reference fraction. Thus squeezing can assist phase-coherent battery energy by suppressing finite-pump number noise, but it cannot replace the pump phase reference itself.

\textit{Cryogenic pump-line implication}.--
The resource distinction identified here has a direct hardware interpretation. In a conventional superconducting readout chain, the phase reference required by a quantum-limited parametric amplifier is supplied by a continuous microwave pump tone. A locally precharged pump battery would be useful only if it stores the phase-coherent component certified above. To estimate the possible scale of the benefit, we use the heat-budget model of Ref.~\cite{Kurman2026}, which assumes one readout-amplifier pump line per eight qubits. In that model, the amplifier pump contributes active heat \(25\,{\rm nW}\) per qubit at the cold plate and \(2.5\,{\rm nW}\) per qubit at the mixing chamber. The passive stainless-steel pump line contributes \(365/8=45.6\,{\rm nW}\) and \(8.5/8=1.06\,{\rm nW}\), respectively. Eliminating or fully duty-cycling this continuous pump line would therefore remove \(\Delta P_{\rm CP}^{\rm pump}\simeq70.6\,{\rm nW}\) and \(\Delta P_{\rm MXC}^{\rm pump}\simeq3.56\,{\rm nW}\) per qubit. For the superconducting-wiring quantum-battery computation architecture of Ref.~\cite{Kurman2026}, whose remaining heat loads are \(P_{\rm CP}=173\,{\rm nW}\) and \(P_{\rm MXC}=5.6\,{\rm nW}\) per qubit, this gives \(P_{\rm CP}\simeq102\,{\rm nW}\) and \(P_{\rm MXC}\simeq2.04\,{\rm nW}\). With cooling powers \(P_{\rm cool,CP}=1000\,\mu{\rm W}\) and \(P_{\rm cool,MXC}=34\,\mu{\rm W}\), the estimate \(N_q^{\max}=\min(P_{\rm cool,CP}/P_{\rm CP},P_{\rm cool,MXC}/P_{\rm MXC})\) changes from \(5.8\times10^3\) to \(9.8\times10^3\), an additional factor of about \(1.7\). This is an architecture-level upper-bound estimate; recharge overhead, residual classical phase calibration, finite pump-battery lifetime, and pump-battery loss would reduce the gain. It nevertheless shows why the resource certified here is experimentally meaningful: replacing a continuous pump tone requires not stored energy alone, but locally stored phase-coherent energy capable of locking the amplifier output. A residual-pump-line version of this estimate is given in Note~4 of the SM \cite{SuppMat}.

\textit{Summary}---
We have shown that a finite quantum pump battery separates two notions that are identified in the classical-pump description of parametric amplification: stored pump energy and phase-coherent work. In the closed trilinear model, coherent and phase-randomized coherent pumps with the same photon-number distribution generate comparable signal-idler populations, but only the coherent pump produces anomalous two-mode coherence and an EPR-squeezed interference dip. The same separation survives in the open system. Cascaded collector modes define genuine temporal output modes in which coherent and phase-randomized pumps deposit nearly the same energy, while only the coherent pump produces sub-vacuum phase-sensitive interference. Thus pair production alone does not certify a useful pump work source; the operational signature is the phase-locked amplifier field.
The squeezed-pump analysis refines this conclusion. Weak amplitude squeezing can slightly improve the temporal-mode interference certifier by reducing finite-pump number fluctuations while preserving most of the coherent displacement. Stronger squeezing reverses this gain because it transfers stored energy away from the first-order phase reference. Squeezing can therefore assist a coherent pump battery by reducing number noise, but it cannot replace the pump phase reference itself. The cryogenic pump-line estimate gives an architecture-level reason why this distinction matters: replacing or strongly duty-cycling a continuous microwave pump tone requires a locally stored phase reference, not stored energy alone. Phase-locked parametric amplification therefore certifies the phase-coherent component of a quantum battery, possibly aided by number-noise reduction, as the part that functions as amplifier work.

\acknowledgments{\emph{Acknowledgments}---BA acknowledges support from IRA Program (project no. FENG.02.01-IP.05-0006/23) financed by the FENG program 2021-2027, Priority FENG.02, Measure FENG.02.01., with the support of the FNP.}

\bibliography{References}

@misc{SuppMat,
  note = {See Supplemental Material for the stiff-pump threshold derivation, cutoff diagnostics, squeezed-pump analysis, and cryogenic pump-line estimate.}
}

@article{AlickiFannes2013,
  author = {Alicki, Robert and Fannes, Mark},
  title = {Entanglement boost for extractable work from ensembles of quantum batteries},
  journal = {Physical Review E},
  volume = {87},
  number = {4},
  pages = {042123},
  year = {2013},
  doi = {10.1103/PhysRevE.87.042123}
}

@article{Kurman2026,
  author = {Kurman, Yaniv and Hymas, Kieran and Fedorov, Arkady and Munro, William J. and Quach, James},
  title = {Powering Quantum Computation with Quantum Batteries},
  journal = {Physical Review X},
  volume = {16},
  pages = {011016},
  year = {2026},
  doi = {10.1103/l39v-jwwz}
}

@article{Krinner2019,
  author = {Krinner, Sebastian and Storz, Simon and Kurpiers, Philipp and Magnard, Paul and Heinsoo, Johannes and Keller, Raphael and Luetolf, Janis and Eichler, Christopher and Wallraff, Andreas},
  title = {Engineering cryogenic setups for 100-qubit scale superconducting circuit systems},
  journal = {EPJ Quantum Technology},
  volume = {6},
  pages = {2},
  year = {2019},
  doi = {10.1140/epjqt/s40507-019-0072-0}
}

@article{Hougland2025,
  author = {Hougland, Nicholas M. and Li, Zhuan and Kaufman, Ryan and Mesits, Boris and Mong, Roger S. K. and Hatridge, Michael and Pekker, David},
  title = {Pump-efficient Josephson parametric amplifiers with high saturation power},
  journal = {Physical Review A},
  volume = {111},
  pages = {022611},
  year = {2025},
  doi = {10.1103/PhysRevA.111.022611}
}

@article{Denney2026,
  author = {Denney, C. and Genter, K. and Cicak, K. and Teufel, J. D. and Aumentado, J. and Lecocq, F. and Malnou, M.},
  title = {A Traveling-Wave Parametric Amplifier With Integrated Diplexers},
  journal = {arXiv:2603.12327},
  year = {2026},
  url={https://doi.org/10.48550/arXiv.2603.12327}
}

@book{MandelWolf1995,
  author = {Mandel, Leonard and Wolf, Emil},
  title = {Optical Coherence and Quantum Optics},
  publisher = {Cambridge University Press},
  address = {Cambridge},
  year = {1995},
  doi = {10.1017/CBO9781139644105}
}

@article{Nehra2026,
  author = {Nehra, N. and Bourlet, N. and Esmaeili, A. H. and Monge, B. and Cyrenne-Bergeron, F. and Paquette, A. and Arabmohammadi, M. and Rogalle, A. and Lapointe, Y. and Hofheinz, M.},
  title = {DC-powered broadband quantum-limited microwave amplifier},
  journal = {Physical Review Applied},
  volume = {25},
  pages = {064009},
  year = {2026},
  doi = {10.1103/c4zz-pbk4}
}

@article{Binder2015,
  author = {Binder, Felix C. and Vinjanampathy, Sai and Modi, Kavan and Goold, John},
  title = {Quantacell: Powerful charging of quantum batteries},
  journal = {New Journal of Physics},
  volume = {17},
  pages = {075015},
  year = {2015},
  doi = {10.1088/1367-2630/17/7/075015}
}

@article{Campaioli2017,
  author = {Campaioli, Francesco and Pollock, Felix A. and Binder, Felix C. and C{\'e}leri, Lucas and Goold, John and Vinjanampathy, Sai and Modi, Kavan},
  title = {Enhancing the Charging Power of Quantum Batteries},
  journal = {Physical Review Letters},
  volume = {118},
  number = {15},
  pages = {150601},
  year = {2017},
  doi = {10.1103/PhysRevLett.118.150601}
}

@article{Ferraro2018,
  author = {Ferraro, Dario and Campisi, Michele and Andolina, Gian Marcello and Pellegrini, Vittorio and Polini, Marco},
  title = {High-Power Collective Charging of a Solid-State Quantum Battery},
  journal = {Physical Review Letters},
  volume = {120},
  number = {11},
  pages = {117702},
  year = {2018},
  doi = {10.1103/PhysRevLett.120.117702}
}

@article{Andolina2018,
  author = {Andolina, Gian Marcello and Farina, Donato and Mari, Andrea and Pellegrini, Vittorio and Giovannetti, Vittorio and Polini, Marco},
  title = {Charger-mediated energy transfer in exactly solvable models for quantum batteries},
  journal = {Physical Review B},
  volume = {98},
  number = {20},
  pages = {205423},
  year = {2018},
  doi = {10.1103/PhysRevB.98.205423}
}

@article{Farina2019,
  author = {Farina, Donato and Andolina, Gian Marcello and Mari, Andrea and Polini, Marco and Giovannetti, Vittorio},
  title = {Charger-mediated energy transfer for quantum batteries: An open-system approach},
  journal = {Physical Review B},
  volume = {99},
  number = {3},
  pages = {035421},
  year = {2019},
  doi = {10.1103/PhysRevB.99.035421}
}

@article{Gherardini2020,
  author = {Gherardini, Stefano and Campaioli, Francesco and Caruso, Filippo and Binder, Felix C.},
  title = {Stabilizing open quantum batteries by sequential measurements},
  journal = {Physical Review Research},
  volume = {2},
  number = {1},
  pages = {013095},
  year = {2020},
  doi = {10.1103/PhysRevResearch.2.013095}
}

@article{Mitchison2021,
  author = {Mitchison, Mark T. and Goold, John and Prior, Javier},
  title = {Charging a quantum battery with linear feedback control},
  journal = {Quantum},
  volume = {5},
  pages = {500},
  year = {2021},
  doi = {10.22331/q-2021-07-13-500}
}

@article{Quach2022,
  author = {Quach, James Q. and McGhee, Kirsty E. and Ganzer, Lucia and Rouse, Dominic M. and Lovett, Brendon W. and Gauger, Erik M. and Keeling, Jonathan and Cerullo, Giulio and Lidzey, David G. and Virgili, Tersilla},
  title = {Superabsorption in an organic microcavity: Toward a quantum battery},
  journal = {Science Advances},
  volume = {8},
  number = {2},
  pages = {eabk3160},
  year = {2022},
  doi = {10.1126/sciadv.abk3160}
}

@article{Lindblad1976,
  author = {Lindblad, G.},
  title = {On the generators of quantum dynamical semigroups},
  journal = {Communications in Mathematical Physics},
  volume = {48},
  pages = {119--130},
  year = {1976},
  doi = {10.1007/BF01608499}
}

@article{GardinerCollett1985,
  author = {Gardiner, C. W. and Collett, M. J.},
  title = {Input and output in damped quantum systems: Quantum stochastic differential equations and the master equation},
  journal = {Physical Review A},
  volume = {31},
  number = {6},
  pages = {3761--3774},
  year = {1985},
  doi = {10.1103/PhysRevA.31.3761}
}

@article{Gardiner1993,
  author = {Gardiner, C. W.},
  title = {Driving a quantum system with the output field from another driven quantum system},
  journal = {Physical Review Letters},
  volume = {70},
  number = {15},
  pages = {2269--2272},
  year = {1993},
  doi = {10.1103/PhysRevLett.70.2269}
}

@article{Carmichael1993,
  author = {Carmichael, H. J.},
  title = {Quantum trajectory theory for cascaded open systems},
  journal = {Physical Review Letters},
  volume = {70},
  number = {15},
  pages = {2273--2276},
  year = {1993},
  doi = {10.1103/PhysRevLett.70.2273}
}

@article{Royer2018,
  author = {Royer, Baptiste and Grimsmo, Arne L. and Choquette-Poitevin, Alexandre and Blais, Alexandre},
  title = {Itinerant Microwave Photon Detector},
  journal = {Physical Review Letters},
  volume = {120},
  number = {20},
  pages = {203602},
  year = {2018},
  doi = {10.1103/PhysRevLett.120.203602}
}

@article{Erdman2024,
  author = {Erdman, Paolo Andrea and Andolina, Gian Marcello and Giovannetti, Vittorio and No{\'e}, Frank},
  title = {Reinforcement Learning Optimization of the Charging of a Dicke Quantum Battery},
  journal = {Physical Review Letters},
  volume = {133},
  number = {24},
  pages = {243602},
  year = {2024},
  doi = {10.1103/PhysRevLett.133.243602}
}

@article{Campaioli2024,
  author = {Campaioli, Francesco and Gherardini, Stefano and Quach, James Q. and Polini, Marco and Andolina, Gian Marcello},
  title = {Colloquium: Quantum batteries},
  journal = {Reviews of Modern Physics},
  volume = {96},
  number = {3},
  pages = {031001},
  year = {2024},
  doi = {10.1103/RevModPhys.96.031001}
}

@article{Caves1982,
  author = {Caves, Carlton M.},
  title = {Quantum limits on noise in linear amplifiers},
  journal = {Physical Review D},
  volume = {26},
  number = {8},
  pages = {1817--1839},
  year = {1982},
  doi = {10.1103/PhysRevD.26.1817}
}

@article{Clerk2010,
  author = {Clerk, A. A. and Devoret, M. H. and Girvin, S. M. and Marquardt, Florian and Schoelkopf, R. J.},
  title = {Introduction to quantum noise, measurement, and amplification},
  journal = {Reviews of Modern Physics},
  volume = {82},
  number = {2},
  pages = {1155--1208},
  year = {2010},
  doi = {10.1103/RevModPhys.82.1155}
}

@article{Bergeal2010,
  author = {Bergeal, N. and Schackert, F. and Metcalfe, M. and Vijay, R. and Manucharyan, V. E. and Frunzio, L. and Prober, D. E. and Schoelkopf, R. J. and Girvin, S. M. and Devoret, M. H.},
  title = {Phase-preserving amplification near the quantum limit with a {Josephson} ring modulator},
  journal = {Nature},
  volume = {465},
  pages = {64--68},
  year = {2010},
  doi = {10.1038/nature09035}
}

@article{WallsBarakat1970,
  author = {Walls, D. F. and Barakat, R.},
  title = {Quantum-Mechanical Amplification and Frequency Conversion with a Trilinear Hamiltonian},
  journal = {Physical Review A},
  volume = {1},
  number = {2},
  pages = {446--451},
  year = {1970},
  doi = {10.1103/PhysRevA.1.446}
}

@article{DrobnyJex1992,
  author = {Drobn{\'y}, Gabriel and Jex, Igor},
  title = {Quantum properties of field modes in trilinear optical processes},
  journal = {Physical Review A},
  volume = {46},
  number = {1},
  pages = {499--506},
  year = {1992},
  doi = {10.1103/PhysRevA.46.499}
}

@article{DrobnyJex1993,
  author = {Drobn{\'y}, Gabriel and Jex, Igor},
  title = {Mode entanglement in nondegenerate down-conversion with quantized pump},
  journal = {Physical Review A},
  volume = {48},
  number = {1},
  pages = {569--579},
  year = {1993},
  doi = {10.1103/PhysRevA.48.569}
}

@article{BandillaDrobnyJex1996,
  author = {Bandilla, A. and Drobn{\'y}, G. and Jex, I.},
  title = {Nondegenerate parametric interactions and nonclassical effects},
  journal = {Physical Review A},
  volume = {53},
  number = {1},
  pages = {507--516},
  year = {1996},
  doi = {10.1103/PhysRevA.53.507}
}

@article{FerneeKinslerDrummond1995,
  author = {Fern{\'e}e, M. and Kinsler, P. and Drummond, P. D.},
  title = {Quadrature squeezing in the nondegenerate parametric amplifier},
  journal = {Physical Review A},
  volume = {51},
  number = {1},
  pages = {864--867},
  year = {1995},
  doi = {10.1103/PhysRevA.51.864}
}

@article{XingRalph2023,
  author = {Xing, Wanli and Ralph, T. C.},
  title = {Pump depletion in optical parametric amplification},
  journal = {Physical Review A},
  volume = {107},
  number = {2},
  pages = {023712},
  year = {2023},
  doi = {10.1103/PhysRevA.107.023712}
}

@article{ChinniEtAl2024,
  author = {Chinni, K. and others},
  title = {Beyond the parametric approximation: Pump depletion, entanglement, and squeezing in macroscopic down-conversion},
  journal = {Physical Review A},
  volume = {110},
  number = {1},
  pages = {013712},
  year = {2024},
  doi = {10.1103/PhysRevA.110.013712}
}

@article{BirrittellaAlsingGerry2020,
  author = {Birrittella, Richard J. and Alsing, Paul M. and Gerry, Christopher C.},
  title = {Phase effects in coherently stimulated down-conversion with a quantized pump field},
  journal = {Physical Review A},
  volume = {101},
  number = {1},
  pages = {013813},
  year = {2020},
  doi = {10.1103/PhysRevA.101.013813}
}

@article{GieseEtAl2018,
  author = {Giese, Enno and Fickler, Robert and Zhang, Wuhong and Chen, Lixiang and Boyd, Robert W.},
  title = {Influence of pump coherence on the quantum properties of spontaneous parametric down-conversion},
  journal = {Physica Scripta},
  volume = {93},
  number = {8},
  pages = {084001},
  year = {2018},
  doi = {10.1088/1402-4896/aace12}
}

@article{ferraro2026opportunities,
  title={Opportunities and challenges of quantum batteries},
  author={Ferraro, Dario and Cavaliere, Fabio and Genoni, Marco G and Benenti, Giuliano and Sassetti, Maura},
  journal={Nature Reviews Physics},
  pages={115–127},
  volume={8},
  year={2026},
  publisher={Nature Publishing Group},
  url={https://doi.org/10.1038/s42254-025-00906-5}
}

@article{PhysRevA.107.042419,
  title = {Catalysis in charging quantum batteries},
  author = {Rodr\'{\i}guez, R. R. and Ahmadi, B. and Mazurek, P. and Barzanjeh, S. and Alicki, R. and Horodecki, P.},
  journal = {Phys. Rev. A},
  volume = {107},
  issue = {4},
  pages = {042419},
  numpages = {8},
  year = {2023},
  month = {Apr},
  publisher = {American Physical Society},
  doi = {10.1103/PhysRevA.107.042419},
  url = {https://link.aps.org/doi/10.1103/PhysRevA.107.042419}
}

@article{ahmadi2026LandauZener,
  title={Quantum-Battery-Powered Geometric Landau-Zener Interferometry},
  author={Ahmadi, Borhan},
  journal={arXiv:2605.18108},
  year={2026},
  url={https://doi.org/10.48550/arXiv.2605.18108}
}

@article{ahmadi2026ChiralSqueezing,
  title={Charging Quantum Batteries with Chiral Squeezing},
  author={Ahmadi, Borhan and H. A. Malavazi, André and Splettstoesser, Janine and Horodecki, Paweł and Du, Lei},
  journal={	arXiv:2606.16764},
  year={2026},
  url={https://doi.org/10.48550/arXiv.2606.16764}
}

@article{PhysRevApplied.23.024010,
  title = {Superoptimal charging of quantum batteries via reservoir engineering: Arbitrary energy transfer unlocked},
  author = {Ahmadi, Borhan and Mazurek, Pawe\l{} and Barzanjeh, Shabir and Horodecki, Pawe\l{}},
  journal = {Phys. Rev. Appl.},
  volume = {23},
  issue = {2},
  pages = {024010},
  numpages = {14},
  year = {2025},
  month = {Feb},
  publisher = {American Physical Society},
  doi = {10.1103/PhysRevApplied.23.024010},
  url = {https://link.aps.org/doi/10.1103/PhysRevApplied.23.024010}
}

@article{PhysRevA.109.042411,
  title = {Better performance of quantum batteries in different environments compared to closed batteries},
  author = {Liu, Shu-Qian and Wang, Lu and Fan, Hao and Wu, Feng-Lin and Liu, Si-Yuan},
  journal = {Phys. Rev. A},
  volume = {109},
  issue = {4},
  pages = {042411},
  numpages = {13},
  year = {2024},
  month = {Apr},
  publisher = {American Physical Society},
  doi = {10.1103/PhysRevA.109.042411},
  url = {https://link.aps.org/doi/10.1103/PhysRevA.109.042411}
}

@article{adma.202415073,
author = {Camposeo, Andrea and Virgili, Tersilla and Lombardi, Floriana and Cerullo, Giulio and Pisignano, Dario and Polini, Marco},
title = {Quantum Batteries: A Materials Science Perspective},
journal = {Advanced Materials},
volume = {37},
number = {17},
pages = {2415073},
doi = {https://doi.org/10.1002/adma.202415073},
url = {https://advanced.onlinelibrary.wiley.com/doi/abs/10.1002/adma.202415073}
}

@article{ahmadi2025harnessing,
  title={Harnessing Environmental Noise for Quantum Energy Storage},
  author={Ahmadi, Borhan and Ravichandran, Aravinth Balaji and Mazurek, Pawe{\l} and Barzanjeh, Shabir and Horodecki, Pawe{\l}},
  journal={arXiv:2510.06384},
  year={2025},
  url={https://doi.org/10.48550/arXiv.2510.06384}
}

@article{yang2024three,
  title={Three-level Dicke quantum battery},
  author={Yang, Dong-Lin and Yang, Fang-Mei and Dou, Fu-Quan},
  journal={Physical Review B},
  volume={109},
  number={23},
  pages={235432},
  year={2024},
  publisher={APS}
}

@article{kamin2023steady,
  title = {Steady-state charging of quantum batteries via dissipative ancillas},
  author = {Kamin, F. H. and Salimi, S. and Arjmandi, M. B.},
  journal = {Phys. Rev. A},
  volume = {109},
  issue = {2},
  pages = {022226},
  numpages = {6},
  year = {2024},
  month = {Feb},
  publisher = {American Physical Society},
  doi = {10.1103/PhysRevA.109.022226},
  url = {https://link.aps.org/doi/10.1103/PhysRevA.109.022226}
}

@article{Razzoli_2025,
doi = {10.1088/2058-9565/ad9ed4},
url = {https://doi.org/10.1088/2058-9565/ad9ed4},
year = {2025},
month = {jan},
publisher = {IOP Publishing},
volume = {10},
number = {1},
pages = {015064},
author = {Razzoli, Luca and Gemme, Giulia and Khomchenko, Ilia and Sassetti, Maura and Ouerdane, Henni and Ferraro, Dario and Benenti, Giuliano},
title = {Cyclic solid-state quantum battery: thermodynamic characterization and quantum hardware simulation},
journal = {Quantum Science and Technology}
}

@article{zakavati2025optimizing,
  title={Optimizing the Charging of Open Quantum Batteries using Long Short-Term Memory-Driven Reinforcement Learning},
  author={Zakavati, Shadab and Salimi, Shahriar and Arash, Behrouz},
  journal={arXiv:2504.19840},
  year={2025},
  url={https://doi.org/10.48550/arXiv.2504.19840}
}

@article{lu2021optimal,
  title = {Optimal state for a Tavis-Cummings quantum battery via the Bethe ansatz method},
  author = {Lu, Wangjun and Chen, Jie and Kuang, Le-Man and Wang, Xiaoguang},
  journal = {Phys. Rev. A},
  volume = {104},
  issue = {4},
  pages = {043706},
  numpages = {13},
  year = {2021},
  month = {Oct},
  publisher = {American Physical Society},
  doi = {10.1103/PhysRevA.104.043706},
  url = {https://link.aps.org/doi/10.1103/PhysRevA.104.043706}
}

@article{qhz8-mvfb,
  title = {Multilevel quantum batteries: Large battery capacity and high charging speed},
  author = {Huang, Zefeng and Zhang, Dayang and Wang, Zhuoheng and Zhao, Yu and Yu, Youbin and Jin, Guangri and Chen, Aixi},
  journal = {Phys. Rev. A},
  volume = {113},
  issue = {4},
  pages = {042616},
  numpages = {7},
  year = {2026},
  month = {Apr},
  publisher = {American Physical Society},
  doi = {10.1103/qhz8-mvfb},
  url = {https://link.aps.org/doi/10.1103/qhz8-mvfb}
}

@article{bv4w-jr6q,
  title = {Two-Time Weak-Measurement Protocol for Ergotropy Protection in Open Quantum Batteries},
  author = {Malavazi, Andr\'e H.A. and Sagar, Rishav and Ahmadi, Borhan and Dieguez, Pedro R.},
  journal = {PRX Energy},
  volume = {4},
  issue = {2},
  pages = {023011},
  numpages = {28},
  year = {2025},
  month = {Jun},
  publisher = {American Physical Society},
  doi = {10.1103/bv4w-jr6q},
  url = {https://link.aps.org/doi/10.1103/bv4w-jr6q}
}

@article{malavazi2025charge,
  title = {Charge-preserving operations in quantum batteries},
  author = {Malavazi, André H. A. and Ahmadi, Borhan and Horodecki, Paweł and Dieguez, Pedro R.},
  journal = {PRX Energy},
  volume = {5},
  pages = {023004},
  year = {2026},
  month = {Mar},
  publisher = {American Physical Society},
  doi = {10.1103/2jtp-jpkn},
  url = {https://doi.org/10.1103/2jtp-jpkn}
}
\clearpage
\onecolumngrid  


\setcounter{section}{0}
\setcounter{subsection}{0}
\setcounter{equation}{0}
\setcounter{figure}{0}
\setcounter{table}{0}

\renewcommand{\theequation}{S\arabic{equation}}
\renewcommand{\thefigure}{S\arabic{figure}}
\renewcommand{\thetable}{S\arabic{table}}

\newcommand{\suppnote}[1]{%
    \refstepcounter{section}%
    \setcounter{subsection}{0}%
    \section*{Supplementary Note \arabic{section}. #1}%
    \addcontentsline{toc}{section}{Supplementary Note \arabic{section}. #1}%
}

\newcommand{\suppsubsection}[1]{%
    \refstepcounter{subsection}%
    \subsection*{\Alph{subsection}. #1}%
    \addcontentsline{toc}{subsection}{\Alph{subsection}. #1}%
}

\section*{Supplementary Materials for "Quantum Batteries as Work Sources for Phase-Locked Parametric Amplification"}

\suppnote{Stiff-pump guide for the phase-coherence threshold}
\label{suppnote:stiff_pump_threshold}

In this note we derive the simple closed-system guide
\begin{equation}
\eta_{c,\rm crit}^{\rm in}
\simeq
\tanh^2\tau ,
\end{equation}
used in the main text to interpret the phase-coherence threshold. This expression is not the exact threshold of the finite quantum-battery model. Rather, it is the undepleted stiff-pump limit, where the pump mode is replaced by a fixed classical amplitude. It provides a useful analytic reference for understanding why higher-gain operation requires a larger phase-coherent fraction of the pump battery.

We start from the trilinear interaction
\begin{equation}
H_I
=
i\hbar g
\left(
a^\dagger b^\dagger c
-
ab c^\dagger
\right),
\end{equation}
where one pump photon in mode \(c\) is converted into one signal-idler pair in modes \(a\) and \(b\). In the stiff-pump limit, the pump is treated as an undepleted phase reference,
\begin{equation}
c\rightarrow \alpha_c ,
\end{equation}
and the interaction reduces to the standard nondegenerate two-mode-squeezing Hamiltonian
\begin{equation}
H_{\rm TMS}
=
i\hbar\lambda
\left(
a^\dagger b^\dagger
-
ab
\right),
\label{eq:SM_TMS_Hamiltonian}
\end{equation}
after choosing the pump phase convention such that \(\lambda=g|\alpha_c|\) is real and positive. We define the dimensionless interaction time
\begin{equation}
\tau=\lambda t .
\end{equation}

The Heisenberg equations generated by Eq.~\eqref{eq:SM_TMS_Hamiltonian} are
\begin{equation}
\frac{d a}{dt}
=
\lambda b^\dagger ,
\qquad
\frac{d b^\dagger}{dt}
=
\lambda a .
\end{equation}
Equivalently,
\begin{equation}
\frac{d}{dt}
\begin{pmatrix}
a\\
b^\dagger
\end{pmatrix}
=
\lambda
\begin{pmatrix}
0 & 1\\
1 & 0
\end{pmatrix}
\begin{pmatrix}
a\\
b^\dagger
\end{pmatrix}.
\end{equation}
Solving these equations gives the Bogoliubov transformation
\begin{equation}
a(t)
=
a(0)\cosh\tau
+
b^\dagger(0)\sinh\tau ,
\label{eq:SM_bog_a}
\end{equation}
and
\begin{equation}
b(t)
=
b(0)\cosh\tau
+
a^\dagger(0)\sinh\tau .
\label{eq:SM_bog_b}
\end{equation}
We assume that the signal and idler modes are initially in vacuum,
\begin{equation}
|\psi(0)\rangle_{ab}
=
|0\rangle_a|0\rangle_b .
\end{equation}
Using Eqs.~\eqref{eq:SM_bog_a} and \eqref{eq:SM_bog_b}, one finds
\begin{equation}
\langle a^\dagger(t)a(t)\rangle
=
\sinh^2\tau ,
\qquad
\langle b^\dagger(t)b(t)\rangle
=
\sinh^2\tau .
\end{equation}
Therefore the generated pair number is
\begin{equation}
n_{\rm pair}
=
\frac{
\langle a^\dagger a\rangle
+
\langle b^\dagger b\rangle
}{2}
=
\sinh^2\tau .
\label{eq:SM_npairs_stiff}
\end{equation}

The anomalous two-mode coherence is
\begin{align}
\langle a(t)b(t)\rangle
&=
\left\langle
\left[
a\cosh\tau+b^\dagger\sinh\tau
\right]
\left[
b\cosh\tau+a^\dagger\sinh\tau
\right]
\right\rangle
\nonumber\\
&=
\cosh^2\tau \langle ab\rangle
+
\sinh\tau\cosh\tau \langle aa^\dagger\rangle
\nonumber\\
&\quad
+
\sinh\tau\cosh\tau \langle b^\dagger b\rangle
+
\sinh^2\tau \langle b^\dagger a^\dagger\rangle .
\end{align}
For the initial vacuum,
\begin{equation}
\langle ab\rangle=0,
\qquad
\langle b^\dagger b\rangle=0,
\qquad
\langle b^\dagger a^\dagger\rangle=0,
\qquad
\langle aa^\dagger\rangle=1 .
\end{equation}
Hence
\begin{equation}
\langle a(t)b(t)\rangle
=
\sinh\tau\cosh\tau .
\label{eq:SM_M_stiff}
\end{equation}
Thus, in the stiff-pump coherent reference,
\begin{equation}
|\langle ab\rangle_{\rm coh}|
=
\sinh\tau\cosh\tau .
\end{equation}

We now connect this result to the phase-coherence threshold. The closed-system phase-sensitive interference observable is
\begin{equation}
I(\phi)
=
\left\langle
\left(a+e^{i\phi}b^\dagger\right)^\dagger
\left(a+e^{i\phi}b^\dagger\right)
\right\rangle .
\end{equation}
Expanding the expression gives
\begin{equation}
I(\phi)
=
n_a+n_b+1
+
2{\rm Re}
\left[
e^{-i\phi}\langle ab\rangle
\right],
\end{equation}
where
\begin{equation}
n_a=\langle a^\dagger a\rangle,
\qquad
n_b=\langle b^\dagger b\rangle .
\end{equation}
The minimum over the analysis phase is therefore
\begin{equation}
I_{\min}
=
n_a+n_b+1
-
2|\langle ab\rangle| .
\label{eq:SM_Imin_closed}
\end{equation}
The vacuum reference is \(I_{\min}=1\). Therefore sub-vacuum phase-sensitive interference requires
\begin{equation}
I_{\min}<1 .
\label{eq:SM_subvacuum_condition}
\end{equation}

For the partially dephased pump family used in the main text,
\begin{equation}
\rho_c(0)
=
w|\alpha\rangle\langle\alpha|
+
(1-w)\rho_{\rm rand},
\end{equation}
the photon-number distribution is unchanged by the dephasing, while the first-order phase coherence is reduced. The initial coherent fraction is
\begin{equation}
\eta_c^{\rm in}
=
\frac{|\langle c\rangle|^2}{\langle c^\dagger c\rangle}
=
w^2 .
\end{equation}
In the stiff-pump guide, this means that the generated population is unchanged, whereas the anomalous coherence is reduced as
\begin{equation}
|\langle ab\rangle|
=
\sqrt{\eta_c^{\rm in}}\,
|\langle ab\rangle_{\rm coh}| .
\label{eq:SM_M_scaling_eta}
\end{equation}
Using \(n_a=n_b=n_{\rm pair}\), Eq.~\eqref{eq:SM_Imin_closed} becomes
\begin{equation}
I_{\min}
=
2n_{\rm pair}
+
1
-
2\sqrt{\eta_c^{\rm in}}\,
|\langle ab\rangle_{\rm coh}| .
\end{equation}
The condition \(I_{\min}<1\) then gives
\begin{equation}
2n_{\rm pair}
+
1
-
2\sqrt{\eta_c^{\rm in}}\,
|\langle ab\rangle_{\rm coh}|
<
1 .
\end{equation}
Canceling the vacuum contribution and dividing by \(2\), we obtain
\begin{equation}
n_{\rm pair}
<
\sqrt{\eta_c^{\rm in}}\,
|\langle ab\rangle_{\rm coh}| .
\end{equation}
Equivalently,
\begin{equation}
\eta_c^{\rm in}
>
\left(
\frac{
n_{\rm pair}
}{
|\langle ab\rangle_{\rm coh}|
}
\right)^2 .
\label{eq:SM_closed_threshold_general}
\end{equation}

Substituting the stiff-pump expressions from Eqs.~\eqref{eq:SM_npairs_stiff} and \eqref{eq:SM_M_stiff} into Eq.~\eqref{eq:SM_closed_threshold_general}, we find
\begin{align}
\eta_{c,\rm crit}^{\rm in}
&\simeq
\left(
\frac{
\sinh^2\tau
}{
\sinh\tau\cosh\tau
}
\right)^2
\nonumber\\
&=
\left(
\frac{\sinh\tau}{\cosh\tau}
\right)^2
\nonumber\\
&=
\tanh^2\tau .
\label{eq:SM_tanh_threshold}
\end{align}
This is the stiff-pump threshold guide quoted in the main text.

The approximation sign in Eq.~\eqref{eq:SM_tanh_threshold} has two origins. First, the derivation assumes an undepleted classical pump, whereas the full model uses a finite pump battery with photon-number-dependent coupling and depletion. Second, the open-system calculations in the main text certify emitted temporal modes, not intracavity equal-time moments. These effects shift the quantitative threshold. Nevertheless, Eq.~\eqref{eq:SM_tanh_threshold} captures the basic mechanism: as the gain parameter \(\tau\) increases, the generated pair population grows relative to the phase-sensitive coherence that can be sustained by a partially coherent pump. Consequently, higher-gain operation requires a larger phase-coherent fraction of the pump battery.

\suppnote{Cutoff convergence of the cascaded collector calculation}
\label{sec:SM_cutoff_convergence}

The open-system results in the main text are obtained from the cascaded collector master equation. The total Hilbert space is truncated as
\begin{equation}
{\cal H}
=
{\cal H}_a
\otimes
{\cal H}_b
\otimes
{\cal H}_c
\otimes
{\cal H}_A
\otimes
{\cal H}_B ,
\end{equation}
with dimensions
\begin{equation}
D
=
N_a N_b N_c N_A^{(f)} N_B^{(f)} .
\end{equation}
Here \(a,b,c\) are the signal, idler, and pump-battery modes, while \(A,B\) are the downstream collector modes used to define the measured temporal output modes. Since the exact density-matrix simulation scales in Liouville space as \(D^2\), a simultaneous large-cutoff sweep in all five modes is computationally very demanding. We therefore perform complementary cutoff checks: one calculation uses larger collector cutoffs, and another uses larger source-mode cutoffs. The operational conclusion is unchanged in both cases.

To diagnose truncation errors, we monitor edge populations. For the source modes we define
\begin{equation}
P_{\rm src}^{\rm edge}(t)
=
\max
\left\{
P_a^{\rm edge}(t),
P_b^{\rm edge}(t),
P_c^{\rm edge}(t)
\right\},
\end{equation}
where, for example,
\begin{equation}
P_a^{\rm edge}(t)
=
{\rm Tr}
\left[
\rho(t)
\,
\left(
|N_a-1\rangle\langle N_a-1|
\right)_a
\right],
\end{equation}
and analogously for \(b\) and \(c\). For the collector modes we define
\begin{equation}
P_{\rm col}^{\rm edge}(t)
=
\max
\left\{
P_A^{\rm edge}(t),
P_B^{\rm edge}(t)
\right\}.
\end{equation}
The collector-mode calculation is performed in a low-excitation parameter regime where the measured temporal modes contain less than one pair on average at the optimal readout time. The Hilbert-space truncation is therefore controlled by monitoring edge populations rather than by increasing all cutoffs uniformly. For the main data set, the source-mode and collector-mode edge populations at the coherent optimum are
\begin{equation}
P_{\rm src}^{\rm edge}(\tau_\star)=1.75\times10^{-3},
\qquad
P_{\rm col}^{\rm edge}(\tau_\star)=7.50\times10^{-3},
\end{equation}
respectively. A complementary calculation with larger source-mode cutoffs gives the same resource hierarchy. Thus the conclusion is not sensitive to the truncation: coherent and phase-randomized pump batteries deposit nearly the same collected energy, while only the coherent pump produces sub-vacuum temporal-mode interference.

We report these quantities both at the physically relevant readout time
\begin{equation}
\tau_\star
=
\arg\min_\tau I_{\min}^{(f)}(\tau)
\end{equation}
for the coherent pump and as a maximum over the simulated time window. The main calculation used in the paper employs
\begin{equation}
N_a=N_b=5,
\qquad
N_c=14,
\qquad
N_A^{(f)}=N_B^{(f)}=5,
\end{equation}
with \(\bar n_c=5\), \(\xi=2\lambda/\sqrt{\kappa_a\kappa_b}=0.8\), \(\gamma_A=\gamma_B=2\lambda\), and \(\lambda T_{\max}=2\). The coherent-pump optimum occurs at
\begin{equation}
\tau_\star = 1.3667 .
\end{equation}
The corresponding cutoff diagnostics are shown in Table~\ref{tab:SM_cutoff_main}.
\begin{table*}
\caption{Cutoff diagnostics for the main cascaded-collector run. All quantities are evaluated at the coherent-pump optimal readout time \(\tau_\star=1.3667\), except for the last two columns, which give the maximum edge populations over the full simulated interval \(0\leq \lambda T\leq 2\). The collector-edge population at \(\tau_\star\) is below \(8\times 10^{-3}\), and the source-edge population is below \(2\times 10^{-3}\).\justifying}
\label{tab:SM_cutoff_main}
\begin{ruledtabular}
\begin{tabular}{lcccccccc}
pump state
&
\(\eta_c^{\rm in}\)
&
\(I_{\min}^{(f)}(\tau_\star)\)
&
\({\cal V}_f(\tau_\star)\)
&
\(C_f(\tau_\star)\)
&
\(n_f(\tau_\star)\)
&
\(P_{\rm src}^{\rm edge}(\tau_\star)\)
&
\(P_{\rm col}^{\rm edge}(\tau_\star)\)
&
\(\max P_{\rm col}^{\rm edge}\)
\\
\hline
coherent
&
0.9966
&
0.5533
&
0.7147
&
0.8341
&
0.4698
&
\(1.75\times10^{-3}\)
&
\(7.50\times10^{-3}\)
&
\(2.52\times10^{-2}\)
\\
partially dephased
&
0.2493
&
1.2459
&
0.3575
&
0.4172
&
0.4695
&
\(1.74\times10^{-3}\)
&
\(7.48\times10^{-3}\)
&
\(2.52\times10^{-2}\)
\\
phase randomized
&
0
&
1.9384
&
0
&
0
&
0.4692
&
\(1.73\times10^{-3}\)
&
\(7.45\times10^{-3}\)
&
\(2.51\times10^{-2}\)
\\
Fock
&
0
&
1.9171
&
0
&
0
&
0.4586
&
\(4.80\times10^{-4}\)
&
\(2.95\times10^{-3}\)
&
\(1.59\times10^{-2}\)
\end{tabular}
\end{ruledtabular}
\end{table*}
The main physical conclusion is insensitive to the cutoff choice. At \(\tau_\star\), the coherent and phase-randomized coherent pumps deposit almost the same energy in the collected modes,
\begin{equation}
n_f^{\rm coh}=0.4698,
\qquad
n_f^{\rm pr}=0.4692,
\end{equation}
but only the coherent pump produces sub-vacuum temporal-mode interference,
\begin{equation}
I_{\min,{\rm coh}}^{(f)}=0.5533<1,
\qquad
I_{\min,{\rm pr}}^{(f)}=1.9384>1 .
\end{equation}
Thus the result is not a consequence of different output energy transfer; it is a consequence of the presence or absence of pump-battery phase coherence.

As a complementary robustness check, we also performed a calculation with larger source-mode cutoffs and slightly smaller collector-mode cutoffs,
\begin{equation}
N_a=N_b=6,
\qquad
N_c=14,
\qquad
N_A^{(f)}=N_B^{(f)}=4 .
\end{equation}
This keeps the exact density-matrix calculation within a stable Liouville-space size while testing the sensitivity to the source-mode truncation. The comparison is shown in Table~\ref{tab:SM_cutoff_robustness}. The numerical values of the certifier change moderately, as expected because the collector cutoff and therefore the collected temporal mode are also changed, but the resource separation is unchanged: the coherent pump gives \(I_{\min}^{(f)}<1\), while the phase-randomized pump remains above the vacuum reference and has nearly identical collected energy.

\begin{table*}[t]
\caption{Complementary cutoff robustness check for the cascaded-collector calculation. The quantities are evaluated at the coherent-pump optimal readout time for each cutoff choice. The main calculation uses larger collector cutoffs, while the second calculation uses larger source-mode cutoffs. In both cases the coherent and phase-randomized coherent pumps deposit nearly the same collected energy, but only the coherent pump gives a sub-vacuum temporal-mode interference dip. Edge populations are reported at the physical readout time \(\tau_\star\), with the maximum collector-edge population over the simulated interval shown for reference.\justifying}
\label{tab:SM_cutoff_robustness}
\begin{ruledtabular}
\begin{tabular}{lcccccccccc}
cutoffs
&
\(D\)
&
\(\tau_\star\)
&
\(I_{\min,{\rm coh}}^{(f)}\)
&
\(I_{\min,{\rm pr}}^{(f)}\)
&
\(C_{f,{\rm coh}}\)
&
\(n_f^{\rm coh}\)
&
\(n_f^{\rm pr}\)
&
\(P_{\rm src}^{\rm edge}(\tau_\star)\)
&
\(P_{\rm col}^{\rm edge}(\tau_\star)\)
&
\(\max P_{\rm col}^{\rm edge}\)
\\
\hline
\((5,5,14,5,5)\)
&
8750
&
1.3667
&
0.5533
&
1.9384
&
0.8341
&
0.4698
&
0.4692
&
\(1.75\times10^{-3}\)
&
\(7.50\times10^{-3}\)
&
\(2.52\times10^{-2}\)
\\
\((6,6,14,4,4)\)
&
8064
&
1.2000
&
0.6178
&
1.7074
&
0.7874
&
0.3541
&
0.3537
&
\(3.92\times10^{-4}\)
&
\(1.57\times10^{-2}\)
&
\(5.80\times10^{-2}\)
\end{tabular}
\end{ruledtabular}
\end{table*}

The comparison in Table~\ref{tab:SM_cutoff_robustness} shows that the central hierarchy of the collector-mode certifier is stable under complementary cutoff choices. The main calculation, \((5,5,14,5,5)\), has the smaller collector-edge population at the physical readout time, while the complementary calculation, \((6,6,14,4,4)\), uses larger source-mode cutoffs. In both cases,
\begin{equation}
n_f^{\rm coh}\simeq n_f^{\rm pr},
\qquad
I_{\min,{\rm coh}}^{(f)}<1,
\qquad
I_{\min,{\rm pr}}^{(f)}>1 .
\end{equation}
The remaining cutoff error is therefore not capable of mimicking the central effect. Increasing the collector cutoff improves the collector-edge diagnostic and deepens the coherent-pump interference dip, while increasing the source cutoffs leaves the same resource separation intact. This supports the use of the \((5,5,14,5,5)\) cutoff set for the main collector-mode figures.

\suppnote{Squeezed quantum battery}
\label{suppnote:squeezed_QB}

In this note we analyze displaced squeezed pump-battery states as a nonclassical stress test of the resource interpretation. The main text compares coherent, phase-randomized coherent, and Fock pump batteries. These states separate stored pump energy from first-order pump phase coherence. Squeezed states provide a complementary test: they can reduce pump-number fluctuations, but at fixed total energy they also redistribute energy away from the coherent displacement that supplies the pump phase reference.

We consider a displaced squeezed pump state
\begin{equation}
|\alpha,\zeta\rangle_c
=
D_c(\alpha)S_c(\zeta)|0\rangle,
\qquad
\zeta=re^{i\vartheta_s},
\end{equation}
where
\begin{equation}
D_c(\alpha)
=
\exp
\left(
\alpha c^\dagger-\alpha^\ast c
\right)
\end{equation}
and
\begin{equation}
S_c(\zeta)
=
\exp
\left[
\frac{1}{2}
\left(
\zeta^\ast c^2-\zeta c^{\dagger 2}
\right)
\right].
\end{equation}
Its mean pump energy is
\begin{equation}
\bar n_c
=
\langle c^\dagger c\rangle
=
|\alpha|^2+\sinh^2 r .
\end{equation}
Thus, when different pump states are compared at fixed stored energy \(\bar n_c\), the displacement amplitude is chosen as
\begin{equation}
|\alpha|^2
=
\bar n_c-\sinh^2 r .
\label{eq:SM_squeezed_fixed_energy_alpha}
\end{equation}
This condition requires \(\bar n_c>\sinh^2 r\).

The first-order coherent fraction of the pump battery is
\begin{equation}
\eta_c^{\rm in}
=
\frac{|\langle c\rangle|^2}{\langle c^\dagger c\rangle}.
\end{equation}
For the displaced squeezed state, \(\langle c\rangle=\alpha\). Therefore, at fixed total energy,
\begin{equation}
\eta_c^{\rm in}
=
\frac{|\alpha|^2}{\bar n_c}
=
1-\frac{\sinh^2 r}{\bar n_c}.
\label{eq:SM_squeezed_eta}
\end{equation}
Equation~\eqref{eq:SM_squeezed_eta} is the central tradeoff. Squeezing can reshape the photon-number distribution, but the energy stored in squeezed fluctuations does not contribute to the first-order pump amplitude that phase-locks the signal-idler output.

For a general squeezed angle, the pump photon-number variance is
\begin{equation}
{\rm Var}(n_c)
=
|\alpha|^2
\left[
\cosh(2r)
-
\sinh(2r)
\cos(2\phi_\alpha-\vartheta_s)
\right]
+
2\sinh^2 r
\left(
\sinh^2 r+1
\right),
\label{eq:SM_squeezed_general_variance}
\end{equation}
where \(\alpha=|\alpha|e^{i\phi_\alpha}\). The amplitude-squeezed choice aligns the squeezed quadrature with the displacement direction,
\begin{equation}
\vartheta_s=2\phi_\alpha .
\end{equation}
Then Eq.~\eqref{eq:SM_squeezed_general_variance} gives
\begin{equation}
{\rm Var}_{\rm amp}(n_c)
=
|\alpha|^2 e^{-2r}
+
2\sinh^2 r
\left(
\sinh^2 r+1
\right).
\label{eq:SM_amp_squeezed_variance}
\end{equation}
By contrast, the phase-squeezed choice,
\begin{equation}
\vartheta_s=2\phi_\alpha+\pi ,
\end{equation}
gives
\begin{equation}
{\rm Var}_{\rm phase}(n_c)
=
|\alpha|^2 e^{2r}
+
2\sinh^2 r
\left(
\sinh^2 r+1
\right).
\label{eq:SM_phase_squeezed_variance}
\end{equation}
Thus amplitude squeezing can reduce pump-number fluctuations, while phase squeezing increases them. However, both reduce the coherent fraction at fixed total energy through Eq.~\eqref{eq:SM_squeezed_eta}.

For weak amplitude squeezing, Eqs.~\eqref{eq:SM_squeezed_fixed_energy_alpha} and \eqref{eq:SM_amp_squeezed_variance} give
\begin{equation}
{\rm Var}_{\rm amp}(n_c)
\simeq
\bar n_c
-
2\bar n_c r
+
(2\bar n_c+1)r^2
+
O(r^3),
\end{equation}
whereas
\begin{equation}
\eta_c^{\rm in}
\simeq
1-\frac{r^2}{\bar n_c}
+
O(r^4).
\end{equation}
Therefore weak amplitude squeezing reduces photon-number fluctuations at first order in \(r\), while the loss of first-order phase coherence starts only at second order. This opens a narrow window in which squeezing can assist the phase-coherent pump battery: it suppresses finite-pump number noise while preserving almost all of the coherent displacement.

The squeezed-vacuum limit makes the distinction especially transparent. For \(\alpha=0\), the pump state has finite stored energy,
\begin{equation}
\bar n_c=\sinh^2 r,
\end{equation}
but
\begin{equation}
\langle c\rangle=0,
\qquad
\eta_c^{\rm in}=0 .
\end{equation}
Moreover, the squeezed vacuum contains only even photon-number components and therefore has no adjacent-number coherence. Since the trilinear interaction converts one pump photon into one signal-idler pair, the phase-sensitive pair amplitude \(\langle A_f B_f\rangle\) requires first-order pump phase coherence. A squeezed-vacuum pump can therefore contain stored energy and can contribute to photon generation, but it cannot generate the first-order anomalous temporal-mode coherence certified by \(I_{\min}^{(f)}<1\).

We now test this tradeoff numerically using the cascaded collector calculation. The parameters are the same as in the main text: \(\xi=2\lambda/\sqrt{\kappa_a\kappa_b}=0.8\), \(\gamma_A=\gamma_B=2\lambda\), and \(\lambda T_{\max}=2\). For each \(\bar n_c\), we compare the coherent pump with the best weakly amplitude-squeezed pump found in the scan. The quantities \(I_{\min}^{(f)}(\tau_\star^{\rm coh})\), \(C_f(\tau_\star^{\rm coh})\), and \(n_f(\tau_\star^{\rm coh})\) are evaluated at the coherent-pump optimal readout time for that value of \(\bar n_c\). We also report the independently optimized minimum \(I_{\min,{\rm best}}^{(f)}\) and the corresponding readout time \(\tau_{\min}\).
\begin{figure*}[t]
\centering
\includegraphics[width=\textwidth]{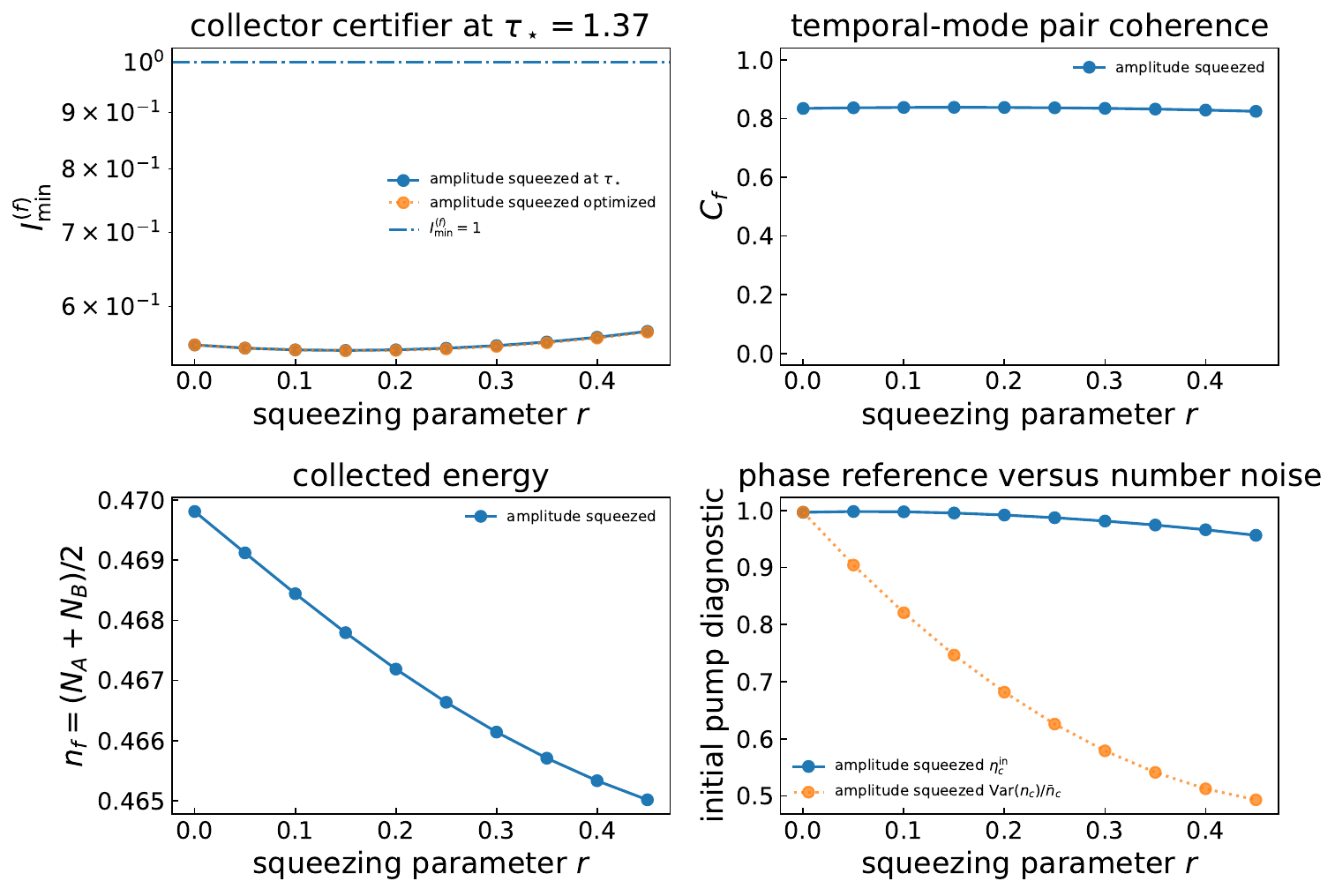}
\caption{Displaced amplitude-squeezed pump batteries at fixed stored energy.
The scan is performed at the main-text operating point \(\bar n_c=5\), \(\xi=0.8\), \(\gamma_A=\gamma_B=2\lambda\), and \(\lambda T_{\max}=2\). The upper-left panel shows the temporal-mode interference certifier \(I_{\min}^{(f)}\) as a function of the squeezing parameter \(r\), both at the coherent-pump optimal readout time and after optimizing over the readout time for each \(r\). Weak amplitude squeezing slightly deepens the sub-vacuum dip, with an optimum near \(r\simeq0.15\). The upper-right and lower-left panels show that the temporal-mode pair coherence \(C_f\) and the collected energy \(n_f\) change only weakly. The lower-right panel shows the underlying tradeoff: amplitude squeezing reduces the pump-number variance while leaving the first-order coherent fraction \(\eta_c^{\rm in}\) close to unity for small \(r\). Thus squeezing can mildly assist a phase-coherent pump battery by reducing finite-pump number noise, but the useful resource remains the phase-coherent displacement.\justifying}
\label{fig:SM_squeezed_scan}
\end{figure*}

The numerical results are summarized in Fig.~\ref{fig:SM_squeezed_scan} and Table~\ref{tab:SM_squeezed_comparison}. Figure~\ref{fig:SM_squeezed_scan} shows the amplitude-squeezed scan at the main-text operating point \(\bar n_c=5\). The temporal-mode interference certifier develops a shallow optimum near \(r\simeq0.15\): weak amplitude squeezing slightly deepens the sub-vacuum dip, while stronger squeezing reverses the gain. The same figure shows why this happens. For small \(r\), the pump-number variance is reduced substantially, whereas the coherent fraction \(\eta_c^{\rm in}\) remains close to unity. Thus the squeezed part of the battery is useful only insofar as it reduces finite-pump number noise without significantly depleting the coherent displacement that phase-locks the amplifier output.

Table~\ref{tab:SM_squeezed_comparison} shows that weak amplitude squeezing provides a small but systematic operational advantage. For \(\bar n_c=2\), the optimized temporal-mode dip improves from
\begin{equation}
I_{\min,{\rm best}}^{(f)}=0.5257
\end{equation}
for the coherent pump to
\begin{equation}
I_{\min,{\rm best}}^{(f)}=0.5156
\end{equation}
for the amplitude-squeezed pump with \(r=0.15\). For \(\bar n_c=5\), the corresponding improvement is
\begin{equation}
0.5533
\rightarrow
0.5466 .
\end{equation}
In both cases the improvement occurs while the coherent fraction remains very close to unity, whereas the normalized pump-number variance is reduced by about $25\%$.

This result refines the main resource statement. Squeezing is not a substitute for phase coherence: a squeezed-vacuum pump has stored energy but no first-order phase reference and cannot produce the anomalous temporal-mode coherence required for \(I_{\min}^{(f)}<1\). Weak amplitude squeezing can nevertheless assist a phase-coherent pump battery by reducing finite-pump number fluctuations without substantially depleting the coherent displacement. Thus the useful resource for phase-preserving amplification is not stored energy alone, nor squeezing alone, but phase-coherent battery energy that can be mildly improved by number-noise reduction.
\begin{table*}[t]
\caption{Weak amplitude-squeezed pump batteries compared with coherent pump batteries at the same stored energy. For both \(\bar n_c=2\) and \(\bar n_c=5\), a small amount of amplitude squeezing improves the temporal-mode interference certifier. The improvement is modest but systematic: squeezing reduces the pump-number variance while keeping the coherent fraction close to unity. The coherent reference at \(\bar n_c=5\) has \(\eta_c^{\rm in}=0.9966\), rather than exactly unity, because of the finite pump Hilbert-space truncation used in the numerical calculation.
\justifying}
\label{tab:SM_squeezed_comparison}
\begin{ruledtabular}
\begin{tabular}{cccccccccc}
\(\bar n_c\)
&
pump state
&
\(r\)
&
\(\eta_c^{\rm in}\)
&
\({\rm Var}(n_c)/\bar n_c\)
&
\(I_{\min}^{(f)}(\tau_\star^{\rm coh})\)
&
\(C_f(\tau_\star^{\rm coh})\)
&
\(n_f(\tau_\star^{\rm coh})\)
&
\(\tau_{\min}\)
&
\(I_{\min,{\rm best}}^{(f)}\)
\\
\hline
2
&
coherent
&
0
&
1.0000
&
1.0000
&
0.5257
&
0.8514
&
0.5079
&
1.5333
&
0.5257
\\
2
&
amplitude squeezed
&
0.15
&
0.9887
&
0.7556
&
0.5195
&
0.8548
&
0.5025
&
1.6667
&
0.5156
\\
5
&
coherent
&
0
&
0.9966
&
0.9969
&
0.5533
&
0.8341
&
0.4698
&
1.3667
&
0.5533
\\
5
&
amplitude squeezed
&
0.15
&
0.9954
&
0.7467
&
0.5470
&
0.8379
&
0.4678
&
1.4000
&
0.5466
\end{tabular}
\end{ruledtabular}
\end{table*}

\suppnote{CRYOGENIC PUMP-LINE ESTIMATE FOR BATTERY-POWERED AMPLIFICATION}

This note gives the heat-load estimate used in the main text. The estimate is based on the cryogenic heat-budget model of Ref.~\cite{Kurman2026}, which itself follows the microwave-line heat-load methodology of Ref.~\cite{Krinner2019}. The purpose is not to claim that the few-photon simulations in the main text already realize a full high-dynamic-range readout amplifier. Rather, the estimate quantifies the possible architecture-level benefit if the continuous microwave pump tone of a phase-preserving amplifier can be replaced, or strongly duty-cycled, by a locally precharged phase-coherent pump battery.

In the model of Ref.~\cite{Kurman2026}, one readout-amplifier pump line serves eight qubits. The active heat associated with this pump line is \(25\,{\rm nW}\) per qubit at the cold plate and \(2.5\,{\rm nW}\) per qubit at the mixing chamber. The passive heat of a stainless-steel microwave pump line is \(365\,{\rm nW}\) at the cold plate and \(8.5\,{\rm nW}\) at the mixing chamber per line. Dividing by eight qubits gives passive contributions \(365/8=45.6\,{\rm nW}\) and \(8.5/8=1.06\,{\rm nW}\) per qubit. Thus, in the ideal pump-line-free limit, the removable heat per qubit is
\begin{equation}
\Delta P_{\rm CP}^{\rm pump}
=
25\,{\rm nW}
+
\frac{365\,{\rm nW}}{8}
=
70.6\,{\rm nW},
\end{equation}
and
\begin{equation}
\Delta P_{\rm MXC}^{\rm pump}
=
2.5\,{\rm nW}
+
\frac{8.5\,{\rm nW}}{8}
=
3.56\,{\rm nW}.
\end{equation}
We also allow for a residual classical pump-line fraction \(\chi\), where \(\chi=1\) corresponds to no pump-line reduction and \(\chi=0\) corresponds to the ideal pump-line-free limit. The heat load at cryogenic stage \(i\in\{{\rm CP},{\rm MXC}\}\) is then
\begin{equation}
P_i(\chi)
=
P_i^{\rm base}
-
(1-\chi)\Delta P_i^{\rm pump}.
\end{equation}
The cooling-limited qubit capacity is estimated as
\begin{equation}
N_q^{\max}(\chi)
=
\min
\left[
\frac{P_{\rm cool,CP}}{P_{\rm CP}(\chi)},
\frac{P_{\rm cool,MXC}}{P_{\rm MXC}(\chi)}
\right],
\end{equation}
with \(P_{\rm cool,CP}=1000\,\mu{\rm W}\) and \(P_{\rm cool,MXC}=34\,\mu{\rm W}\). The resulting ideal pump-line-free estimates are summarized in Table~\ref{tab:SM_pump_line_capacity}.

\begin{table*}[t]
\caption{
Cryogenic heat-load estimate for replacing the continuous readout-amplifier pump tone by a locally precharged phase-coherent pump battery. The baseline heat loads are taken from the superconducting-wiring architectures of Ref.~\cite{Kurman2026}. We assume one amplifier pump line per eight qubits, with removable active pump-line heat \(25\,{\rm nW}\) per qubit at the cold plate and \(2.5\,{\rm nW}\) per qubit at the mixing chamber, and removable passive stainless-steel pump-line heat \(365/8\,{\rm nW}\) and \(8.5/8\,{\rm nW}\), respectively. The table shows the ideal pump-line-free limit \(\chi=0\). The capacity gain is reported relative to the corresponding architecture without the battery-powered amplifier pump, namely relative to the first row for the standard superconducting-wiring architecture and relative to the third row for the quantum-battery-computation architecture.\justifying
}
\label{tab:SM_pump_line_capacity}
\begin{ruledtabular}
\begin{tabular}{lcccc}
architecture
&
\(P_q^{\rm CP}\)
&
\(P_q^{\rm MXC}\)
&
\(N_q^{\max}\)
&
capacity gain
\\
\hline
standard with superconducting wiring
&
\(701\,{\rm nW}\)
&
\(15.2\,{\rm nW}\)
&
1426
&
1
\\
standard with superconducting wiring and battery-powered amplifier pump
&
\(630\,{\rm nW}\)
&
\(11.6\,{\rm nW}\)
&
1586
&
1.11
\\
QB computation with superconducting wiring
&
\(173\,{\rm nW}\)
&
\(5.6\,{\rm nW}\)
&
5755
&
1
\\
QB computation with superconducting wiring and battery-powered amplifier pump
&
\(102\,{\rm nW}\)
&
\(2.04\,{\rm nW}\)
&
\(9.8\times10^3\)
&
1.70
\end{tabular}
\end{ruledtabular}
\end{table*}

Table~\ref{tab:SM_pump_line_capacity} should therefore be read as two paired comparisons. For the standard superconducting-wiring architecture, removing the continuous amplifier pump line changes the cooling-limited capacity from \(1426\) to \(1586\), corresponding to a modest gain of \(1.11\). For the superconducting-wiring quantum-battery-computation architecture of Ref.~\cite{Kurman2026}, the same pump-line removal changes the heat loads from \(P_q^{\rm CP}=173\,{\rm nW}\) and \(P_q^{\rm MXC}=5.6\,{\rm nW}\) to \(P_q^{\rm CP}\simeq102\,{\rm nW}\) and \(P_q^{\rm MXC}\simeq2.04\,{\rm nW}\). The resulting cooling-limited capacity increases from \(5755\) to \(9.8\times10^3\), an additional factor of \(1.70\) relative to that architecture without the battery-powered amplifier pump.

This estimate should be interpreted as an upper bound on the architectural benefit. It assumes that the continuous amplifier pump tone can be replaced by a locally precharged pump mode without adding an equivalent heat load during the measurement window. In a realistic implementation, recharge overhead, finite pump-battery lifetime, residual phase calibration, and nonideal pump-battery efficiency would reduce the gain. The estimate nevertheless identifies why the resource certified in the main text is relevant to scalable readout: a pump battery intended to replace a classical parametric-amplifier pump must store phase-coherent energy, not merely energy.

\end{document}